\documentclass[]{aa}  
\usepackage{graphicx,natbib}
\usepackage{epsfig,epstopdf}
\usepackage{ulem}
\usepackage[]{inputenc} %latin1
\usepackage[T1]{fontenc}
\bibpunct{(}{)}{;}{a}{}{,}
\usepackage{txfonts}
\include{def}
\usepackage{changebar}
\usepackage{booktabs}
\usepackage{caption}
\usepackage{ctable}
\usepackage{mathrsfs}

\begin{document}

   \title{Critical core mass for enriched envelopes: the role of H$_2$O condensation}

    \author{J. Venturini \inst{1}  
       \and
       Y. Alibert \inst{1, 2}
       \and
       W. Benz \inst{1}
       \and
       M. Ikoma \inst{3}
     }
\offprints{J. Venturini \\ \email julia.venturini@space.unibe.ch}
\institute{Center for Space and Habitability \& Physikalisches Institut, Universitaet Bern, CH-3012 Bern, Switzerland
\and
Observatoire de Besan\c con, 41 avenue de l'Observatoire, 25000 Besan\c con, France
\and
Department of Earth and Planetary Science, The University
of Tokyo, 7-3-1 Hongo, Bunkyo-ku, Tokyo 113-0033, Japan}

  %the planetesimals are accreted by protoplanets without undergoing any physical disruption during their way to the core.

   \date{\today}

% \abstract{}{}{}{}{} 
% 5 {} token are mandatory
 
  \abstract
    {Within the core accretion scenario of planetary formation, most simulations performed so far always assume the accreting envelope to have a solar composition. From the study of meteorite showers on Earth and numerical simulations, we know that planetesimals must undergo thermal ablation and disruption when crossing a protoplanetary envelope. Thus, once the protoplanet has acquired an atmosphere, not all planetesimals reach the core intact, i.e., the primordial envelope (mainly H and He) gets enriched in volatiles and silicates from the planetesimals. This change of envelope composition during the formation can have a significant effect in the final atmospheric composition and on the formation timescale of giant planets.}
   {To investigate the physical implications of considering the envelope enrichment of protoplanets due to the disruption of icy planetesimals during their way to the core. Particular focus is placed on the effect on the critical core mass for envelopes where condensation of water can occur.}
   {Internal structure models are numerically solved with the implementation of updated opacities for all ranges of metallicities and the software CEA to compute the equation of state. CEA package computes the chemical equilibrium for an arbitrary mixture of gases and allows the condensation of some species, including water. This means that the latent heat of phase transitions is consistently incorporated in the total energy budget.}
   {The critical core mass is found to decrease significantly when an enriched envelope composition is considered in the internal structure equations. A particular strong reduction of the critical core mass is obtained for planets whose envelope metallicity is larger than $Z \approx$ 0.45 when the outer boundary conditions are suitable for condensation of water to occur in the top layers of the atmosphere. We show that this effect is qualitatively preserved also when the atmosphere is out of chemical equilibrium.}
   {Our results indicate that the effect of water condensation in the envelope of protoplanets can severely affect the critical core mass, and should be considered in future studies.}


   \maketitle

\section{Introduction}\label{intro}
The metallicity of the giant planets of our solar system shows that they are considerably enriched in heavy elements compared to solar composition \citep[][and references therein]{GuillRev}. Even though this change of composition with respect to solar could be a consequence of an eroding core \citep{2004jpsm.book...35G,2012ApJ...745...54W}, numerical studies have shown that the change of composition induced by the disruption of planetesimals in the envelope of growing giant planets can be considerable \citep{1988Icar...73..163P,2006tafp.conf...84M, 2007Icar..187..600I,2013ApJ...775...80F}. For instance, a protoplanet with a core of 6 $\text{M}_{\oplus}$ can have an envelope dense enough to completely destroy icy planetesimals of up to 100 km before they reach the core \citep{2006A&A...450.1221B}. The dominant size of planetesimals at the time planet formation takes place might even be smaller, as is suggested by studies which incorporate the collisional fragmentation in the evolution of planetesimals \citep{
2003Icar..166...46I, 2011ApJ...738...35K, 
2014arXiv1401.7738G}. This means that a considerable enrichment can readily occur even for protoplanets with small cores. Thereby, the contribution of metals (compounds heavier than H and He) to the total mass of the envelope of a protogiant planet starts to be relevant already in the early stages of growth. We show in this paper that taking into account this early enrichment of the envelope can have significant consequence on the subsequent growth and evolution of a planet. 

The aforementioned studies \citep{1988Icar...73..163P,2006tafp.conf...84M, 2007Icar..187..600I} have focused on computing the trajectory of planetesimals during their way to the core, as well as their mass and energy deposition in the envelope due to thermal ablation and disruption. However, the resultant metallicity was not included in the thermochemical properties of the envelope, i.e., it was not taken into account to solve the internal structure equations. 

The opposite focus was given by \citet{2011MNRAS.416.1419H} (hereafter HI11). In the first part of their work, they assumed a given envelope metallicity and incorporated a consistent equation of state and opacity tables to solve the static internal structure equations. ``Static'' means that the only source of energy considered is the release of gravitational energy due to the accretion of planetesimals. This implies that the energy release due to the contraction of the envelope (which necessarily provides a varying luminosity with time) is neglected.

Static internal structure calculations allow for the computation of the critical core mass, which corresponds to the maximum mass of the core able to sustain an envelope in hydrostatic equilibrium \citep{1974Icar...22..416P,1978PThPh..60..699M,1980PThPh..64..544M}. Analytical work by  \citet{1982P&SS...30..755S} and \citet{1993Icar..106..323W} (for fully radiative and fully convective envelopes, respectively) show that when the core becomes critical, the mass of the envelope is of the same order of magnitude than the core mass. This means that at the critical point the self-gravity of the envelope becomes important. Further, quasi-static studies of planet growth  \citep{1986Icar...67..391B, 1996Icar..124...62P}, show that when the energy release due to the contraction of the envelope is taken into account, rapid gas accretion takes place when $M_{\text{core}} \approx M_{\text{env}}$, that is, when the core is around criticality. Since the timescale of gas accretion at critical mass is much shorter 
than 
the one required to build a critical core, reaching criticality is a synonym of being able to form a giant planet. Hence, the relevance of the critical core mass concept in the core-accretion scenario.

The enrichment of the envelope due to planetesimal disruption causes a reduction of the critical core mass due to the increase in mean molecular weight of the envelope \citep[first pointed out by ][]{1982P&SS...30..755S}, but more importantly, due to the reduction of the adiabatic temperature gradient that takes place when chemical reactions among volatile molecules occur (HI11). In this work we have included the effect of water latent heat and studied its effect on the critical core mass.

The role played by the condensation of water (i.e., the formation of clouds) in Earth's atmosphere is well known among the climate community \citep[see, e.g.,][chap.~2]{2010ppc..book.....P}. When vapor raises above the condensation level, water droplets form and in the phase transition process latent heat is released. This modifies the energy budget of the atmosphere, causing the pressure-temperature profile to switch from \textit{dry} to \textit{moist} adiabat. We will show that this change in the temperature profile actually can lead to a severe reduction of the critical core mass.

The paper is organized as follows. The basic equations, assumptions and numerical scheme are described in Sect. \ref{numerics}. The opacities and equation of state implemented are described in Sect. \ref{opacities} and \ref{EOS}, respectively. Sect. \ref{resultados} is devoted to the results concerning  the effect  on the critical core mass due to water condensation in an envelope of uniform metallicity. In Sect. \ref{disc} we justify our assumptions concerning the use of an ideal equation of state and uniform metallicity. We discuss as well the dependence of our results on the boundary conditions, and the implications of our results on the diversity of planets that can be formed. Finally, Sect. \ref{conclusions} summarises our main conclusions.

\section{Numerical approach}\label{numerics}
We have developed a code that solves the static internal structure equations assuming spherical symmetry and a constant luminosity (or constant accretion rate of solids when specified). For better numerical performance, the equations are written as a function of $v = r^3 / M_{\text{r}}$, where $r$ is the radial coordinate of the planet, and $M_{\text{r}}$ the mass enclosed inside a sphere of radius $r$:

\begin{equation} \label{hydeq}
 \frac{dP}{dv}= - G \rho_{gas} \bigg(\frac{M_{\text{r}}}{v^2}\bigg)^{2/3} [3 - 4 \pi \rho_{tot} v]^{-1} 
\end{equation}
\begin{equation} \label{mass_cons}
 \frac{dM_{\text{r}}}{dv}= 4\pi \rho_{tot} M_{\text{r}} [3 - 4 \pi \rho_{tot} v]^{-1}
\end{equation}
\begin{equation}\label{transport}
 \frac{dT}{dv}=\frac{T}{P}\frac{dP}{dv}\nabla
\end{equation}\\
where $P$ is the pressure, $\rho_{tot}$ the total density and $\rho_{gas}$ the gas density \footnote{$\rho_{tot}$ includes the mass of all species and $\rho_{gas}$ just the one of the gaseous component. These two densities are equal if no condensation takes place. See Sect. \ref{condensed_species} for more details.}, $T$ the temperature, and $G$ the gravitational constant. The gradient in Eq.(\ref{transport}) is defined as $\nabla \equiv \big(\frac{\partial \text{ln} T}{\partial \text{ln} P}\big)$. For the radiative case it is equal to:
\begin{equation}\label{rad_grad}
 \nabla_{\text{rad}}= \frac{3 \kappa L P}{64 \pi \sigma G M_{\text{r}} T^4}
\end{equation}
$\sigma$ being the Stefan–Boltzmann constant, $\kappa$ the opacity and $L$ the luminosity. For the convective transport $\nabla$ is the gradient evaluated at constant entropy (adiabatic gradient):
\begin{equation}\label{ad_grad}
 \nabla_{\text{ad}}=\bigg(\frac{\partial \text{ln} T}{\partial \text{\text{ln}} P}\bigg)_{\text{S}}
\end{equation}
Convection occurs if $ \nabla_{\text{ad}} < \nabla_{\text{rad}}$ (Schwarzschild criterion).
\bigskip

The three differential equations are solved using a 4th- order Runge-Kutta algorithm. Two boundary conditions are specified at the top of the gaseous envelope ($T_{\text{out}}$ and $P_{\text{out}}$) and one at the core (core mass or core density). The outer boundary conditions and the position of the planet are listed in Table \ref{tabCB}. The total mass of the planet ($M_{\text{P}}$) is defined as the mass inside the Bondi or Hill radius, whichever is smaller. They are, respectively:
\begin{equation}\label{Rp}
 R_B = \frac{GM_{\text{P}}}{c_s^2},\\
 R_H = \bigg(\frac{M_{\text{P}}}{3 M_{\odot}}\bigg)^{1/3} a
\end{equation}\\
where $c_s$ is the sound speed and $a$ the semimajor axis of the planet.

Two different schemes to solve the above equations were implemented. In the first approach (used for the results shown in Fig.\ref{increasing_Z}), the core mass ($M_{\text{core}}$), the core density ($\rho_{\text{core}}$), and the accretion rate of solids ($\dot{M}_{\text{core}}$) are given. Then the total mass of the planet ($M_{\text{P}}$) is computed through an iterative process. A first value for the total mass is assumed and the structure equations are solved starting from the outer boundary and stopping at the core. The total mass is iterated until the mass of the core is reached within a given tolerance. Since in this approach the core radius ($R_{\text{core}}$) and core mass ($M_{\text{core}}$) are known a priori, the luminosity of the envelope is given by the release of the gravitational potential energy of the planetesimals:
\begin{equation}\label{Mdotcore}
 L =\frac{G M_{\text{core}} \dot{M}_{\text{core}}}{R_{\text{core}}}
\end{equation}
which implies that the energy is assumed to be deposited at the surface of the core. This is not consistent with our assumption of the disruption of the planetesimals occurring before they reach the core, but to do this properly, a varying luminosity in the envelope should be implemented, computed from the mass and energy deposition of the disrupted planetesimals. This will be done in a future work. However, for the cases of highly enriched envelopes (the focus of this work), the envelopes are fully convective (see Sect.\ref{uniformZ}), and hence, the thermal structure is independent on $L$. Thus, the precise value of $L$ is not expected to affect our results.  

In the second approach, the total mass of the planet and total luminosity are given. $\dot{M}_{\text{core}}$ is not known a priori because $ M_{\text{core}}$ is not known. In this case the integration is also performed from the surface to the core, but the difference is that now $M_{\text{P}}$ is known, so no iteration is needed. The integration stops when $\rho_r$ ($\rho_r=\frac{3 M_r}{4 \pi r^3}$), is equal to the density of the core. The mass inside this radius is then defined as $M_{\text{core}}$. The independent variable $v$ in Eqs.\ref{hydeq}, \ref{mass_cons}, \ref{transport} was introduced with the purpose of making it possible for the integration to stop exactly at $\rho_r = \rho_{\text{core}}$. Otherwise, an interpolation in $r$ would be required to obtain the mass at which  $\rho_r = \rho_{\text{core}}$. This last method is the one used in all the results shown in this work, except for those shown in Figs. \ref{mmw_Z_profiles}, \ref{ad_grad_Z_profiles} and \ref{increasing_Z}.
Its advantage resides in the fact that no iterations are needed to obtain $M_{\text{core}}$, which makes the scheme more stable for core masses close to the critical core mass.

\begin{table}
\centering
\begin{tabular}{ | l | c |}
\hline\
  $a $ & 5.2 AU \\
  $T_{\text{out}}$ & 150 K \\
  $P_{\text{out}}$  & 0.267 dyn/cm$^2$  \\
  $\rho_{\text{core}}$ & 3.2 g/cm$^3$\\
  L & $1.0 \times 10^{27}$ erg/s\\
\hline  
\end{tabular}
\caption{Parameters used in all simulations. The planet location and boundary conditions defined here are referred in the main text as ``standard case''. These parameters are only modified in Sect. \ref{test_CB}.}
\label{tabCB}
\end{table} 

\section{Opacities}\label{opacities}
Most of the opacity tables available in the literature are for solar metallicity. In order to study the effect of enrichment of the envelopes of protoplanets in a consistent way, we must use opacity tables for all ranges of metallicities. In this work we implement, as in HI11, gas opacity tables for all metallicities computed by J. Ferguson \citep[based on the calculations of][]{1994ApJ...437..879A} and Rosseland mean opacities of dust calculated by \citet{2003A&A...410..611S}.

The Ferguson opacities assume scaled solar abundances (the relative abundances of the various elements is assumed solar but the total abundance is scaled). The range of validity of the tables is for $T \ge 1000$ K and  $-2 \le$ log R $\le 6$, where $R = \rho/T_6^3$, $\rho$ being the gas density and $T_6$ the temperature in million kelvin. Whenever the envelope values of temperature and density lay outside these ranges, the tables are extrapolated. Using a linear or constant extrapolation does not affect our results.

As in HI11, we define the total opacity as:
\begin{equation}\label{kappa}
 \kappa = \kappa_{\text{gas}} + f \kappa_{\text{dust}}
\end{equation}
where $\kappa_{\text{gas}}$ and $\kappa_{\text{dust}}$ are the Rosseland mean opacity of gas and dust grains, respectively. The reduction/enhancement factor $f$ includes the unknown values of the size of the dust grains in protoplanetary envelopes. In principle, the grains released by the ablation of planetesimals could increase the dust opacity with respect to the disc values, but also the grains could rapidly coagulate and hence increase their size, reducing in this way the opacity of the dust \citep{2003Icar..165..428P, 2008ASPC..398..257M, 2010Icar..209..616M, 2014arXiv1406.4127M, 2014ApJ...789L..18O}.

In this work we adopt values of $f=1$, since the effect of other reduction/enhancement values was already investigated by HI11. Using $f=10$ hardly affects the critical core values with respect of using $f=1$, and using lower values of $f$ reduces the critical core mass even further. Thereby, setting $f=1$ provides upper values for the critical core mass as far as the effect of dust opacity is concerned. Regardless of the fact that these are upper limits, we will show that they become increasingly smaller with increasing metallicity. 

In Sect.\ref{remarks_opac} we make use of the \citet{2008ApJS..174..504F} opacities, which are detailed computed opacity tables for H and He and solar metallicity. We use these opacities to illustrate the effect of different opacity tables in the case of envelopes with extended radiative regions. All other results were computed making use of the opacity given by Eq.\ref{kappa} .

\section{Equation of state}\label{EOS}

Beyond suitable opacities, an enriched envelope needs also to be described with a proper equation of state. It turns out that, to the best of our knowledge, there is no equation of state available for an arbitrary metallicity which takes also into account the degeneracy pressure of free electrons.  The latter effect play an important role in the more massive objects. One EOS accounting for these effects is the \citet{1995ApJS...99..713S} EOS for a mixture of H and He (hereafter SCVH). 

In this work we make use of the publicly available software CEA (Chemical Equilibrium with Applications) developed by NASA \citep{nla.cat-vn4108353}. This program solves the chemical equilibrium equations for an arbitrary defined gas mixture. It assumes an ideal gas equation of state for the mixture, but takes into account the proper dissociation and ionization temperatures of each compound formed. Despite the fact that CEA does not consider the pressure of degenerate electrons (which translates into an overestimate of central temperature and pressure for massive objects), we have used it to study the effect of condensation on the envelope structure.

CEA has the advantage of containing a database of $\sim1000$ compounds, and the fact that it considers condensed species, like H$_2$O in solid and liquid phase.\footnote{For the complete list of condensed species considered in CEA, see Appendix B of the User's Manual from http://www.grc.nasa.gov/WWW/CEAWeb/ceaWhat.htm}  The latent heat of phase transitions is included in the calculation of the thermodynamical quantities. For an explanation on the calculation of the density when condensation takes place, see Sect.\ref{condensed_species}. For a discussion on the condensation of other species than water, see Sect. \ref{ammonia}.

The program can be customized to use as an input the pressure, temperature and composition in molar or mass fraction.
In our simulations, the envelope is assumed to be made of hydrogen, helium, carbon and oxygen. We take a fixed C/O mass ratio of 0.24:0.69, which is the one corresponding to Comet Halley \citep{1993prpl.conf.1177M}, value that we assume representative of the early solar system. The metallicity, $Z$, is defined as: 
\begin{center}
 \begin{equation}
Z = \frac{M_{\text{Z, env}}}{M_{\text{env}}}
\end{equation}
\end{center}
where $M_{\text{Z, env}}$ is the mass of elements heavier than H and He in the envelope, and $M_{\text{env}}$ the total mass of the envelope.

Carbon and oxygen constitute the Z-component of the envelope. $Z$ is a free parameter in our study, and is assumed to be uniform throughout the envelope. This assumption will be justified in Sect. \ref{disc}. The mass abundance of H and He ($X$ and $Y$, respectively) are computed as a weighted average between the solar and Comet Halley values, as in HI11. \footnote{This choice of planetesimal composition is dictated by the comparison with earlier work of HI11. These ratios could be varied to match, e.g, the composition of different stars.}

CEA gives as an output the density, mean molecular weight, internal energy, specific heat, and logarithmic derivatives of certain thermodynamical quantities from which it is possible to compute the adiabatic gradient. 

The adiabatic gradient (defined by eq. \ref{ad_grad}) is an important quantity because, together with the radiative gradient (eq. \ref{rad_grad}), it determines the type of heat transport present in the envelope. Given its relevance to our study of the structure of planetary envelopes, we present in the next section how we determine this gradient. 

\subsection{Adiabatic gradient with CEA}
One of the outputs of CEA is the log-derivative of the volume with respect to temperature at constant pressure. From this quantity and the specific heat at constant pressure, the adiabatic gradient can be easily calculated.

From the implicit function theorem:
\begin{equation}\label{implicit}
 \bigg(\frac{\partial T}{\partial P} \bigg)_{\text{S}} \bigg(\frac{\partial P}{\partial S} \bigg)_T  \bigg(\frac{\partial S}{\partial T} \bigg)_P = -1
\end{equation}

From the definition of the adiabatic gradient, 
\begin{equation}\label{dTdPs}
 \nabla_{\text{ad}} = \frac{P}{T} \bigg(\frac{\partial T}{\partial P}\bigg)_{\text{S}}
\end{equation}
Then,
\begin{equation}
 \nabla_{\text{ad}} =  -\frac{P}{T}\frac{(\partial S/\partial P)_T}{(\partial S/\partial T)_P}
\end{equation}
From the definition of the heat capacity at constant pressure ($C_P$):
\begin{equation}\label{cp}
 \bigg(\frac{\partial S}{\partial T} \bigg)_P =\frac{C_P}{T}
\end{equation}
Deriving the Gibbs free energy ($G = U -TS +PV$) with respect to $T$ and $P$ it can be shown that:
\begin{equation}
 \bigg(\frac{\partial S}{\partial P} \bigg)_T = - \bigg(\frac{\partial V}{\partial T} \bigg)_P
\end{equation}
And finally, substituting by the log-derivatives and using the fact that the mixture is an ideal gas:
\begin{equation}
  \nabla_{\text{ad}} = \frac{nR}{C_P} \bigg(\frac{\partial \text{\text{ln}} V}{\partial \text{ln} T}\bigg)_P 
\end{equation}
where $n$ is the number of gas moles and $R$ the gas constant.

\subsection{Condensed species in CEA: density calculation}\label{condensed_species}
In CEA the mean molecular weight is defined as $\mathcal{M}= \frac{M_{tot}}{n_{gas}}$ ($M_{tot}$ being the total mass of the system and $n_{gas}$ the number of moles of the gas), which is not the mean molecular weight of the gas in the case when condensation occurs.

It can be shown that the molecular weight of the gas satisfies: 
\begin{equation}\label{mu_gas}
 \mathcal{M}_{gas} = (1 - X_c)\mathcal{M} 
\end{equation}
where $X_c$ is the mass fraction of the condensed species, quantity readily computed with CEA using Eqs. 2.3b and 2.4a of \citet{nla.cat-vn4108353}. Since the density computed by the program satisfies: 
\begin{equation}
 \rho_{tot} = \frac{\mathcal{M} P}{RT}
\end{equation}
And the gas density must satisfy: \footnote{or equivalently, $\rho_{gas}$ =  $\frac{\mu_{gas} m_A P}{k_B T}$, being $m_A$ the atomic mass unit and $k_B$ the Boltzmann constant. The mean molecular weight of the gas, $\mu_{gas}$, is numerically equivalent to $\mathcal{M}_{gas}$ but is unitless.}
\begin{equation}
 \rho_{gas} = \frac{\mathcal{M}_{gas} P}{RT}
\end{equation}
The gas density is readily obtained from: 
\begin{equation}
 \rho_{gas} = (1 - X_c)\rho_{tot} 
\end{equation}
This density enters in the hydrostatic equilibrium equation, because the pressure gradient is exerted just by the gas.\footnote{It is debated in the literature if actually $\rho_{gas}$ or $ \rho_{tot}$ should be used in the hydrostatic equilibrium equation, but recent work suggest that the correct expression is with $\rho_{gas}$ \citep{2012JETP..115..723G}. In any case, we have checked that using $\rho_{gas}$ or $\rho_{tot}$ in the hydrostatic equilibrium equation does not affect our results.} For the mass conservation equation, the density that includes the total mass, i.e. $\rho_{tot}$, is the one that must be used. (See Sect.\ref{numerics}).

\section{Critical core mass for different uniform metallicities}\label{resultados}

\subsection{Comparison with HI11}\label{compHI}

The static internal structure equations were solved for a range of uniform envelope metallicities between 0.1 and 0.95, and a wide range in protoplanetary mass. The planet location, boundary conditions and envelope composition were taken as in HI11 (see Table \ref{tabCB}), in order to be able to perform comparison tests. The equality in composition means that we set the same $X$ and $Y$ for a given $Z$ (see Sect.\ref{EOS}), and that we allowed the same 13 species in gaseous phase (H, He, O, C, $\text{H}_{\text{2}}$, $\text{O}_{\text{2}}$, $\text{CO}_{\text{2}}$, $\text{H}_{\text{2}} \text{O}$, $\text{CH}_{\text{4}}$, $\text{H}^{\text{+}}$, $\text{O}^{\text{-}}$ and $\text{e}^{\text{-}}$) to be formed when chemical equilibrium is computed. Hereafter we will refer to this case as the ``restricted'' case.

\begin{figure}
\begin{center}
\includegraphics[width=\columnwidth]{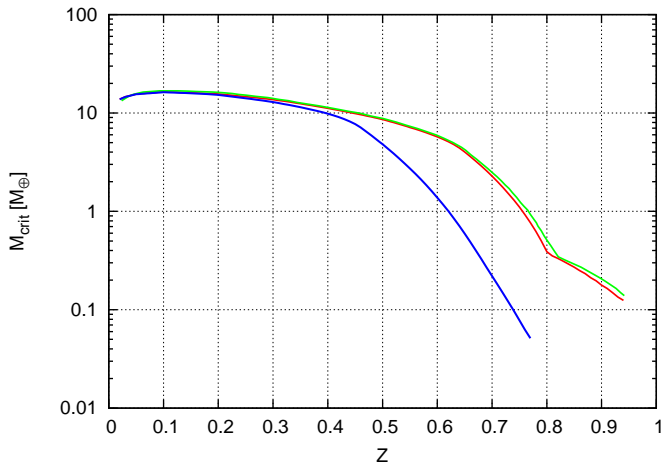}
   \caption{Critical core mass as a function of metallicity. Green line: results of the case ``wholly polluted'' of Fig.2 of HI11, red line: this work, restricted case. Blue line: this work, non-restricted case. (See main text for the definitions of the different cases).}
    \label{Mcrit_Z}
\end{center}
\end{figure}

A comparison with HI11 is shown in Fig.\ref{Mcrit_Z} (green versus red curve). The agreement between the two calculations is remarkable, especially considering that a different way of computing the EOS was used in the different works (HI11 did not use CEA, but computed the thermodynamical quantities partly from the NIST tables, see HI11).

\subsection{Inclusion of $\text{H}_{\text{2}} \text{O}$ condensation} \label{results}
The blue line of Fig.\ref{Mcrit_Z} shows the results for the same setup, but in this case no restriction is imposed in the compounds that can be formed when the chemical equilibrium is computed. Hence, CEA makes use of its chemical database, and allows $\sim$ 50 compounds to be formed from the same proportion of H, He, C and O than before. It also allows condensed species to be formed. We will refer to this case, where no restriction is imposed in the species that can be formed at equilibrium, the ``non-restricted'' case.

Regardless of the case, the most remarkable feature shown by Fig.\ref{Mcrit_Z} is the drastical reduction of the critical core mass with increasing Z. Comparing now the ``non-restricted'' with the ``restricted'' case, it can be seen that the reduction of $M_{\text{crit}}$ with increasing Z is larger for the first case. Namely, for $Z$ = 0.7, the difference in critical core mass between the restricted and the non-restricted case, is of $\Delta M_{\text{crit}}\approx 2 \text{M}_{\oplus}$, and since the critical mass of the non-restricted case is $M_{\text{crit,NR}} \approx 0.2 \text{M}_{\oplus}$, the relative difference is $\frac{\Delta M_{\text{crit}}}{M_{\text{crit,NR}}}  \approx 10$.  

\begin{figure}
\begin{center}
\includegraphics[width=\columnwidth]{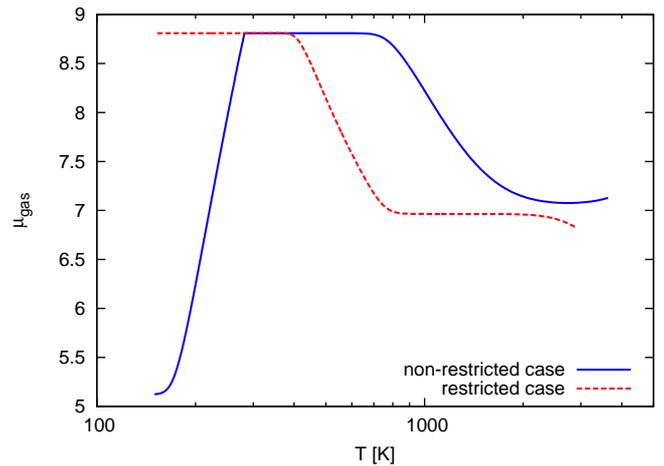} 
   \caption{Profile of the gas mean molecular weight of the envelope as a function of its temperature for $Z$ = 0.7 when $M_{\text{core}} = M_{\text{crit,NR}} (Z=0.7)$. Solid lines: non-restricted case. Dotted lines: restricted case. Note that in the the non-restricted case, $\mu_{gas}$ decreases in the outer layers (i.e., low temperature values) when comparing to the restricted case. This effect is explained in the main text. The difference between both profiles at high temperatures is a consequence of a difference in the pressure-temperature profiles, see Fig. \ref{Z07_Pvapor}.}
    \label{mmw_Z_profiles}
\end{center}
\end{figure}

\begin{figure}
\begin{center}
\includegraphics[width=\columnwidth]{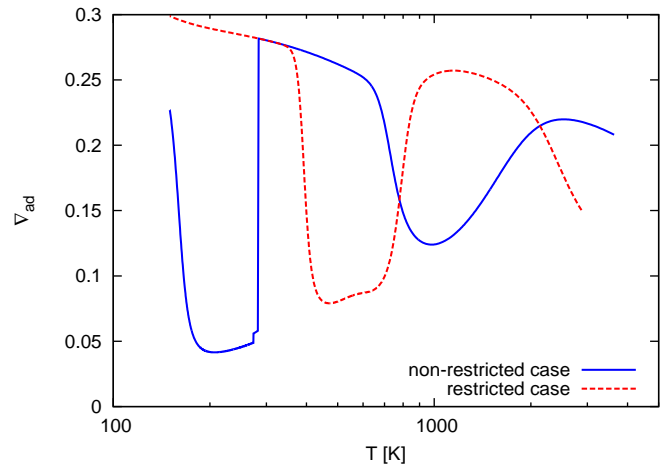}
   \caption{Profile of the adiabatic gradient of the envelope as a function of its temperature for $Z$ = 0.7. The core mass and labels are set just as in Fig. \ref{mmw_Z_profiles}. Note that there is a strong reduction of the adiabatic gradient in the outer layers of the envelope (i.e., low temperature values) in the non-restricted case. This effect is explained in the main text. The difference between both profiles at high temperatures has the same explanation as in the previous figure.}
   \label{ad_grad_Z_profiles}                                                                  
\end{center}
\end{figure}

\begin{figure*}  
\centering
\begin{tabular}{c c}
\includegraphics[width=0.5\textwidth]{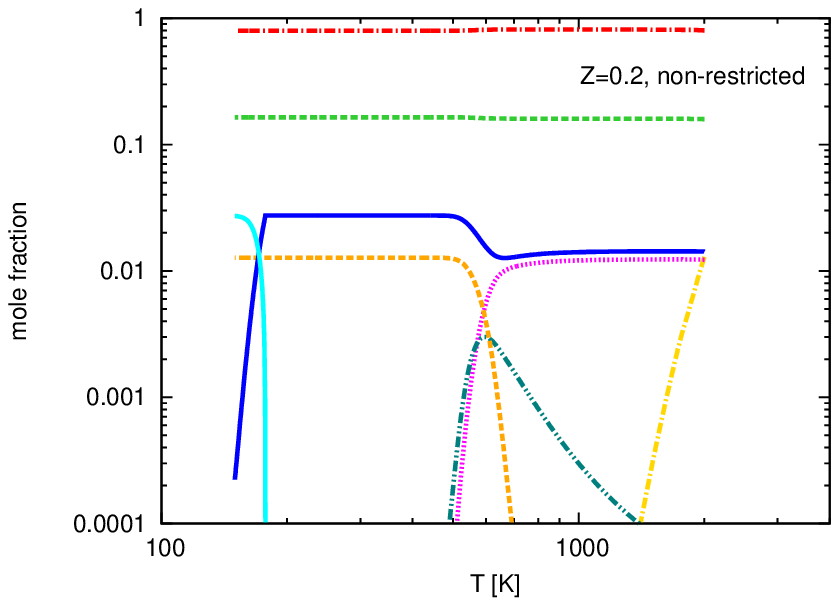} 
\includegraphics[width=0.5\textwidth]{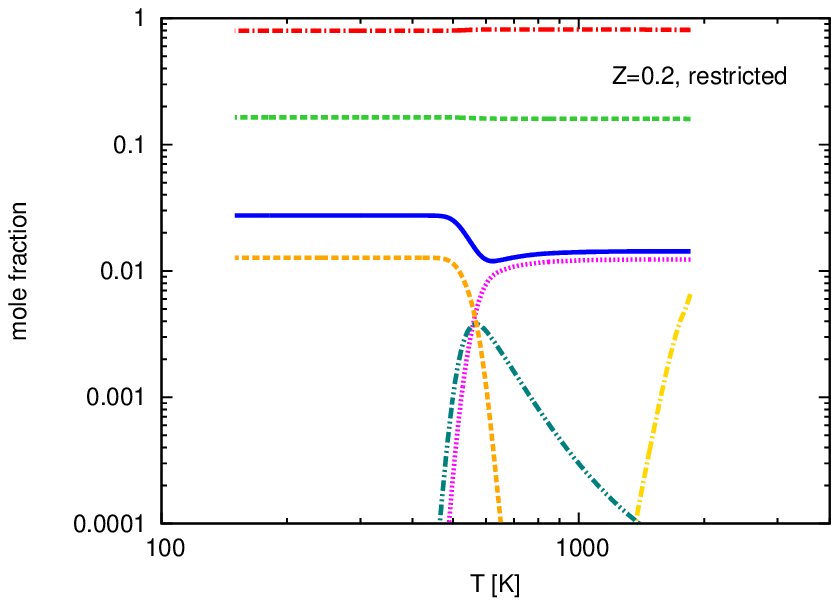} \\
\includegraphics[width=0.5\textwidth]{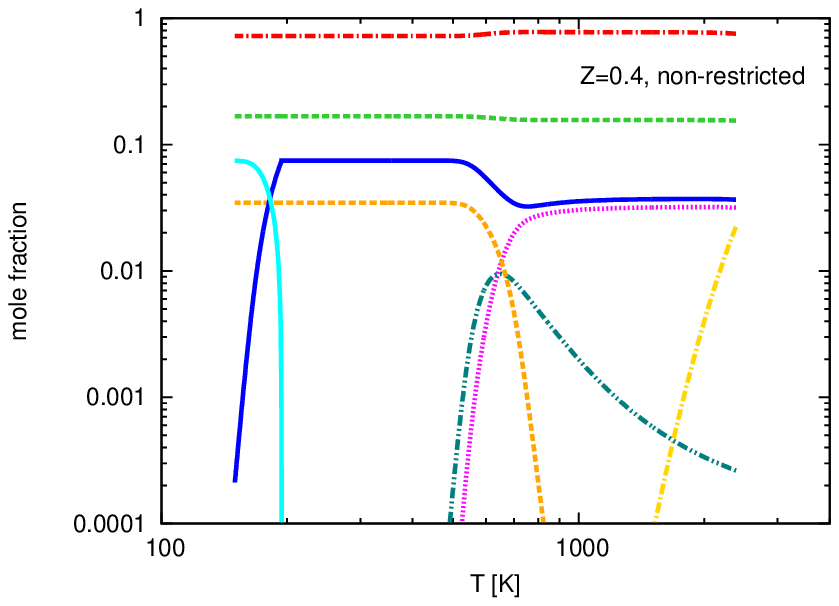} 
\includegraphics[width=0.5\textwidth]{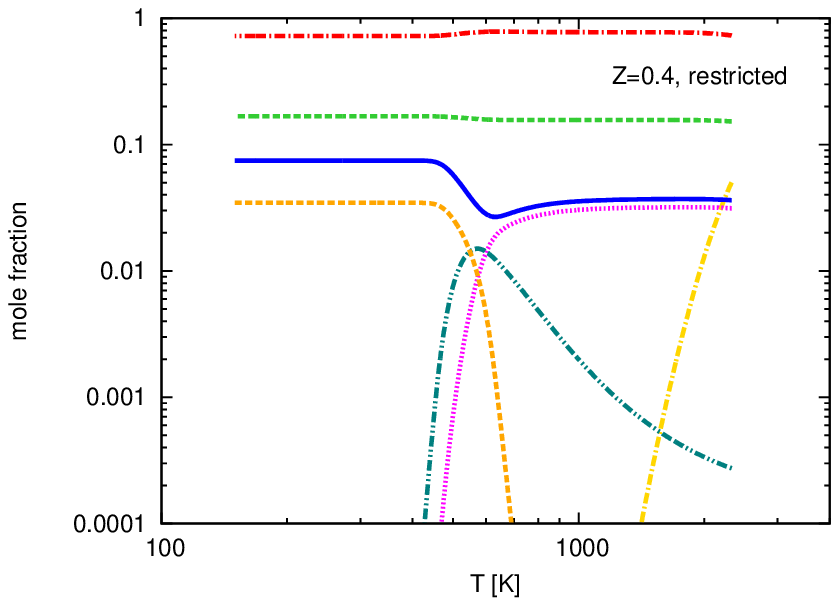} \\
\includegraphics[width=0.5\textwidth]{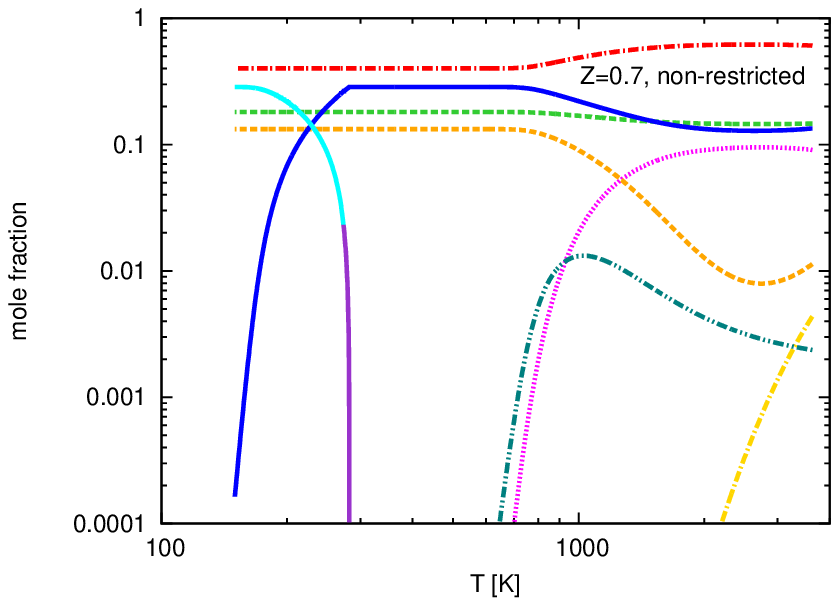}
\includegraphics[width=0.5\textwidth]{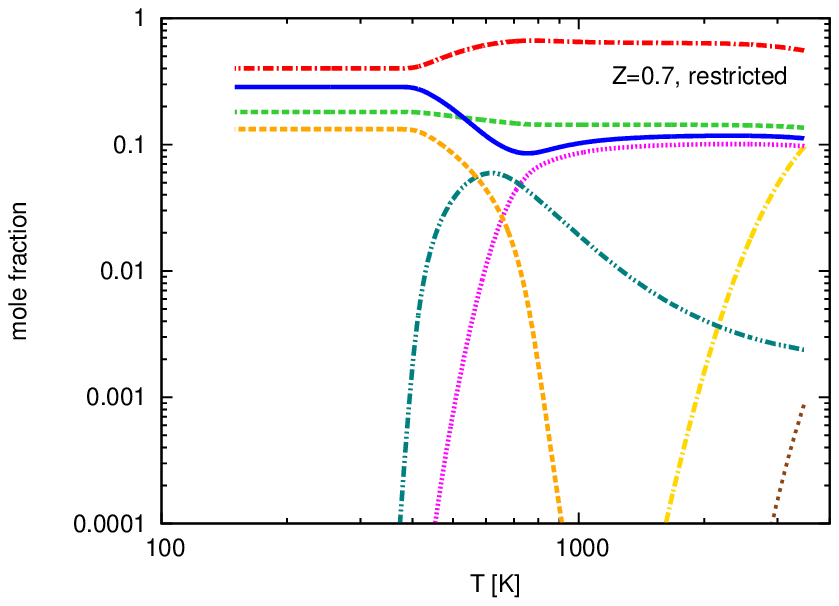} 
\end{tabular}
\includegraphics[width=0.8\textwidth]{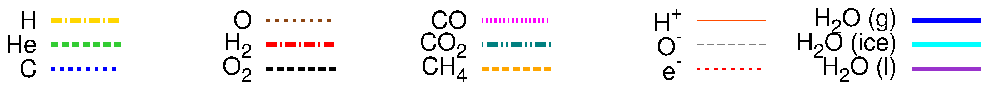} \\
\caption{Envelope composition as a function of temperature for $Z$ = 0.2, 0.4, and 0.7. $M_{\text{P}}$=0.29 $\text{M}_{\oplus}$ for all profiles (planetary mass when $M_{\text{core}}$ (Z=0.7) = $M_{\text{crit}}$ (Z=0.7)). Left column: non-restricted case. Right column: restricted case. In the non-resticted case it can be appreciated that water (solid lines) coexist in vapor phase (blue curve), solid phase (cyan curve) and even with liquid phase (violet curve) in the case of $Z$ = 0.7. Except H$_2$O in solid and liquid phase, the other 13 species shown here were readily considered in HI11. In the restricted case, water only exists in vapor phase.}
\label{compos}
\end{figure*}

In order to understand the difference in $M_{\text{crit}}$ between the ``restricted'' and ``non-restricted'' cases, we plot the gas mean molecular weight and adiabatic gradient as a function of the envelope temperature for $Z$ = 0.7. For both profiles we assume the same core mass, which corresponds to the critical core of the non-restricted case for $Z$ = 0.7. The purpose of this choice is to ensure the same core gravity. This, in turn, ensures that the differences observed in the atmospheric profiles originate from the chemistry and not from a difference in initial pressure gradients. 
The results are shown in Figs. \ref{mmw_Z_profiles} and \ref{ad_grad_Z_profiles}. A decrease of the gas mean molecular weight and of the adiabatic gradient in the outer layers are observed when comparing the non-restricted with the restricted case (solid and dotted lines, respectively). By analyzing the composition in the outer layers of the envelope of the non-restricted case, we find that the 13 species assumed by HI11 are indeed the dominating ones in the CEA calculations as well, as shows Fig. \ref{compos} for $Z = 0.7$. \footnote{The envelope composition for $Z$ = 0.2 and 0.4 are also shown in Fig. \ref{compos}, to illustrate how the abundance of volatiles, in particular water, increases with increasing metallicity.} However, we also find that large amounts of water are being condensed in the coldest layers of the envelope of the non-restricted case, fact which does not occur in the restricted case (see Fig. \ref{compos}). In fact, observing Figs. \ref{mmw_Z_profiles} and \ref{compos}, it can 
be appreciated that the remarkable difference in mean molecular weight in the outer part of the envelope between the non-restricted and restricted cases lasts as long as H$_2$O coexist in both vapor and solid phase. This is a consequence of what was stated in Sect.\ref{condensed_species}: the larger the mass fraction of condensed species, the lower $\mu_{gas}$ (see Eq.\ref{mu_gas}). 

Water condensation in the outer layers of the envelope modifies the molecular weight and adiabatic gradient. In order to assess which of the two effects plays a role in reducing $M_{\text{crit}}$, we performed the following test. We computed the envelope profile for $Z=0.7$ and $M_{\text{P}}$ = 0.29 $\text{M}_{\oplus}$ using $\mu$ of the restricted case and $\nabla_{\text{ad}}$ of the non-restricted case. The profile follows exactly the same profile as the non-restricted case. Therefore, the reduction of the adiabatic gradient is what causes the smaller critical mass.

A straightforward way to understand why the reduction of the adiabatic gradient tends to lower the critical core mass is as follows: given a fixed core mass (as in the cases shown in Fig.\ref{ad_grad_Z_profiles}), the gravity is fixed, and therefore the pressure gradient (Eq. \ref{hydeq}) is fixed. The thermal profile is established by Eq.(\ref{transport}). 
From the ideal gas equation we have:
\begin{equation}
 {P} = \frac{\rho k T}{\mu m_A}
\end{equation}
(where $\mu$ is the mean molecular weight of the gas mixture and $m_A$ the atomic mass unit). 
If the envelope is convective and $\nabla_{\text{ad}}$ diminishes, the temperature for a given radius diminishes. Hence, in order to keep the same pressure gradient, the density of the envelope has to increase. Since the envelope mass is the integral of the density, if the density increases, the envelope mass increases as well. The more massive an envelope gets for a given core mass, the more likely it is that the core becomes critical (see Fig.\ref{PTcurves}).

\begin{figure}  
\begin{center}
\includegraphics[width=\columnwidth]{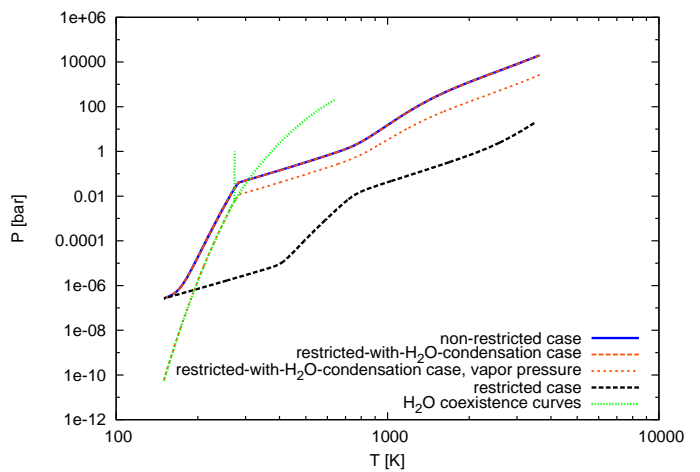}
   \caption{Pressure as a function of temperature for Z=0.7 and $M_{\text{P}}=0.29$ $\text{M}_{\oplus}$. In the case where water is allowed to condense (orange lines), the vapor pressure follows, until T=282 K, the Clausius-Clapeyron curve, which defines the loci where the vapor and liquid phase, and vapor and ice phase, can co-exist (green-dotted line). Hence, in these range of temperatures, water is being condensed, and the pressure-temperature profiles follow the ``moist adiabat".}
   \label{Z07_Pvapor}
\end{center}
\end{figure}

\begin{figure}  
\begin{center}
\includegraphics[width=\columnwidth]{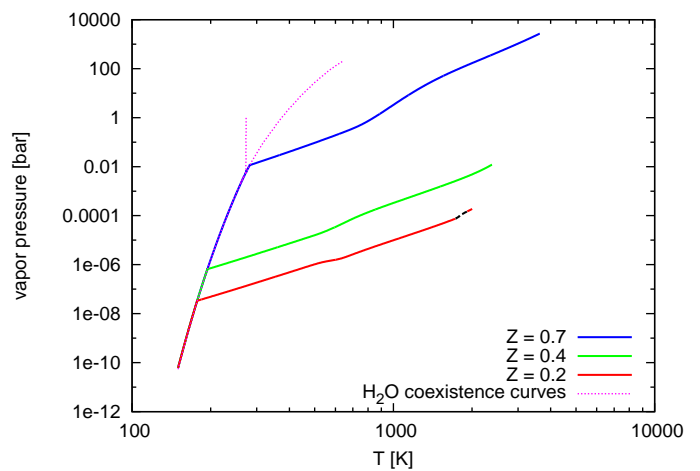}
   \caption{Pressure as a function of temperature for Z=0.2, 0.4 and 0.7. $M_{\text{P}}=0.29$ $\text{M}_{\oplus}$ for all cases. The dotted-black lines correspond to regions where the envelope is radiative. This region is extremely tiny in the case of Z=0.2 profile.}
   \label{Pvapor_Z_varias}
\end{center}
\end{figure}

As stated before, the formation of other compounds in the non-restricted case does not play a role, since the 13 compounds assumed by HI11 are the most abundant ones when chemical equilibrium is computed. But taking into account the condensation really changes the picture, as illustrated in Fig.\ref{Z07_Pvapor}. In this figure, the pressure-temperature profile for Z=0.7 for the non-restricted case,  and the restricted case without condensation of water, are shown. Superposed, we plot the case where the composition is ``restricted'', but where water is allowed to condense (hereafter we will refer to this case as “restricted-with-water-condensation”). The vapor pressure for this last case is also plotted. From this figure it is clear that $\text{H}_{\text{2}} \text{O}$ is being condensed from the top of the atmosphere until T=282 K, since in this range of temperature the vapor pressure is equal to the saturation pressure of water (green- dotted line, given by the Clausius-Clapeyron relation). As the temperature increases from the top of the envelope to the surface of the core, there is a location from which water can only exist in vapor phase. For the mass abundance of water when $Z$ = 0.7, this happens at T=282 K. If the abundance of water is lower than in the $Z$ = 0.7 case, the shift between ``moist'' and ``dry'' adiabat \citep[as is usually called in atmospheric studies, see, e.g.,][chap.~2]{2010ppc..book.....P} occurs for lower temperatures (see Fig.\ref{Pvapor_Z_varias}). This is because the vapor pressure is proportional to the mole fraction of vapor (law of partial pressures). This depends on the amount of water present in the outer layers of the envelope, which increases with metallicity, as shows Fig.\ref{compos}. 

Returning to Fig.\ref{Z07_Pvapor}, it can be noticed that due to water condensation, $\big(\frac{dT}{dP}\big)_{moist}$ < $\big(\frac{dT}{dP}\big)_{dry}$ in the coldest layers of the atmosphere. This is the cause of the strong reduction in the adiabatic gradient of the non-restricted case shown in Fig.\ref{ad_grad_Z_profiles}. 

Fig.\ref{Z07_Pvapor} can also be used to understand why condensation of water reduces $M_{\text{crit}}$. If we look at the envelope profile from the surface of the core to the top of the atmosphere, when we compare the non-restricted with the restricted case, we see that at the pressure (equivalent to atmospheric height) where water starts to condense, the temperature of the non-restricted case does not drop so much as in the restricted case. Since latent heat is released in the transition vapor-solid and vapor-liquid, the envelope gets warmer when condensation of water takes place. The release of latent heat is what really makes the difference between the ``moist'' and ``dry'' profiles. It modifies the pressure-temperature profile remarkably, making it possible to have a much higher pressure for a given temperature, which translates in a much higher envelope density, and therefore, in a considerable reduction of critical core mass.

\subsection{Increasing envelope metallicity}\label{enriching}
\begin{figure}
\begin{center}
\includegraphics[width=\columnwidth]{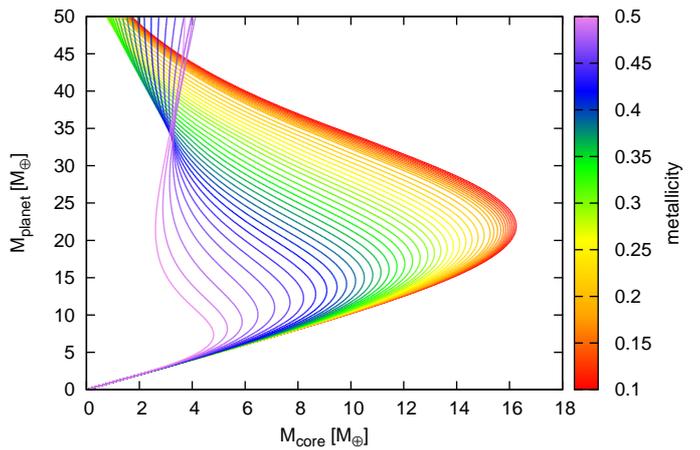}
   \caption{Total mass of the planet as a function of the mass of the core for different uniform metallicities and $L=10^{27}$ erg/s. Here the composition of the envelope corresponds to the ``non-restricted case'' (see main text). The highest $Z$ shown is 0.5 for aesthetic reasons.}
    \label{PTcurves}
\end{center}
\end{figure}

In the previous section,  we confirmed an earlier finding of \citet{2011MNRAS.416.1419H} that an enrichment of the envelope leads to a reduction of the mass of the core at which gas can be accreted in a runaway fashion.  In addition, we pointed out that in case the enrichment leads to condensation, the reduction of the critical core mass is even greater (see Figs.\ref{Mcrit_Z} and \ref{PTcurves}) and can actually become the dominant effect.

To pose this enrichment problem in an evolutionary context, it is interesting to see what would be the total mass of a planet that becomes critical from a gradual enriching process, and how enriched the envelope becomes when reaching the critical core mass.
In order to tackle this, we performed the following calculation. We assumed that a core grows up to 6 $\text{M}_{\oplus}$ due to the accretion of planetesimals that reach the core without being destroyed in the envelope. In other words, all the planetesimals reach the core  without undergoing any physical disruption or thermal ablation until the core acquires a mass of 6 $\text{M}_{\oplus}$. Once $M_{\text{core}}$ = 6 $\text{M}_{\oplus}$, we assume that the protoplanet has acquired an atmosphere sufficiently massive to completely disrupt the infalling planetesimals before they reach the core. This choice of 6 $\text{M}_{\oplus}$ as a threshold is completely arbitrary. It corresponds to the case of icy planetesimals with a radius of 100 km or less \citep{2006A&A...450.1221B}. However, as we explained in Sect. \ref{intro}, planetesimals could be smaller. In that case, full disruption could already occur for planets with a smaller core; a case also analysed at the end of this section.

\begin{figure}
\includegraphics[width=\columnwidth]{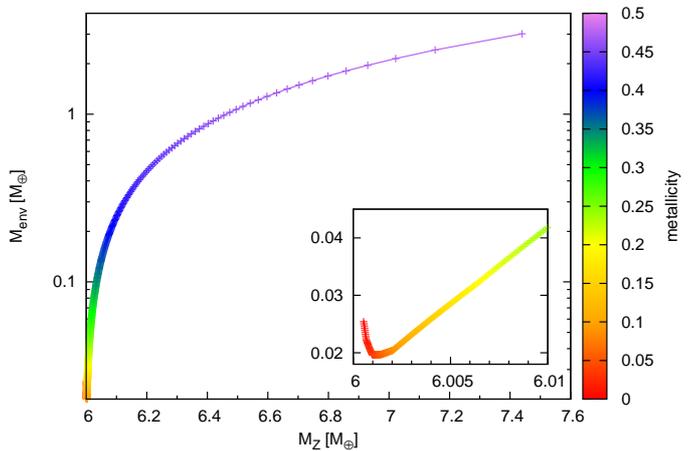}
   \caption{Envelope mass as a function of the mass of solids contained in the planet ($M_Z$). The mass of the core is fixed at 6 $\text{M}_{\oplus}$. An accretion rate of solids of $3\times 10^{-6}$ $\text{M}_{\oplus}$/yr was assumed, which correspond to a luminosity of $10^{27}$ erg/s (the same used in all the other results). The decrease of $M_{\text{env}}$ for low metallicities (shown in the insert) is due to the increase in the gas opacity and the fact that for these cases of low Z, a radiative zone develops in the outer layers. For Z $>$ 0.1, the structures are almost fully convective. Therefore the increase in $\kappa_{\text{gas}}$ does not play a role, the dominant effect is the increase in $M_{\text{env}}$ due to the inclusion of Z in the EOS.}
    \label{increasing_Z}
\end{figure}

For the case of a fixed core mass of 6 $\text{M}_{\oplus}$ as the core threshold for completely disrupting planetesimals in the envelope, \footnote{Since we need to fix the core mass at the beginning of the simulations, we use, in this case, the first numerical scheme described in Sect.\ref{numerics}.} we solved the internal structure equations, and let the envelope metallicity grow gradually until the planet becomes critical. The results are shown in Fig. \ref{increasing_Z}, where the total mass of the envelope as a function of the total mass of the heavy elements acquired by the planet is plotted. If initially, when the fully disruption of the planetesimals starts, $Z$ = 0.02 ($Z_{\odot}$), then the envelope is able to accrete planetesimals until reaching a metallicity of $\approx 0.48$, which represents an envelope mass content of heavy elements of 1.4 $\text{M}_{\oplus}$. At this point the core becomes critical, i.e., no more solution to the static internal equations can be found. According to the 
standard view of the core-accretion model, the subsequent evolution of the 
protoplanet would imply the accretion of large amounts of gas in a runaway fashion \citep{1980PThPh..64..544M, 1986Icar...67..391B}, making it possible for the protoplanet to become a giant planet soon after the protoplanet becomes critical.
The results found here suggest that after the disruption of planetesimals in the envelope begins, the planetary envelope does not need to grow much in order for the planet to become critical. In the particular case of the simulations shown here, a protoplanet with $M_{\text{core}}$ = 6 $\text{M}_{\oplus}$ needs to accrete 1.4 $\text{M}_{\oplus}$ of H and He and the same amount of heavy elements to become critical. A consistent evolutionary study, where the metallicity is computed from the accretion rate of solids, will be done in the future to be able to account for the timescales to trigger runaway of gas when the envelope enrichment is taken into account. 

A similar analysis of the enrichment process can also be drawn for cores smaller than 6 $\text{M}_{\oplus}$ by just making use of Fig.\ref{Mcrit_Z}. If we assume, for instance, that the core stops to grow at 1 $\text{M}_{\oplus}$,  then the envelope will get gradually enriched until the core becomes critical at Z $\approx$ 0.62 (see blue curve of Fig.\ref{Mcrit_Z}). The corresponding mass of the planet in this critical case is $M_{\text{P}}$ = 1.5 $\text{M}_{\oplus}$, which means that in this case the mass of heavy elements in the envelope when $M_{\text{crit}}$ is reached would be $\approx$ 0.3 $\text{M}_{\oplus}$, and the mass of H-He  $\approx$ 0.2. $\text{M}_{\oplus}$. It would be interesting also for cases like this, where $M_{\text{crit}}$ is low and the envelopes are heavily polluted, to study the subsequent evolution. The fate of these small but highly enriched critical protoplanets is not easy to assess \textit{a priori}, since the gravity of the core could not be large enough to trigger gas runaway \citep{2000ApJ...537.1013I, 2010ApJ...714.1343H}. In principle, if runaway of gas accretion cannot be reached, the planet would continue its growth by slowly contracting (and therefore, also slowly) accreting gas. If runaway of gas is not triggered during the lifetime of the gas disc, the result could be the formation of a super-Earth, heavily enriched planet, which would be a very interesting scenario, given the increasing amount of Earth and super-Earth sized planets that are being found \citep{2013ApJ...770...69P}, and the inference of a very enriched composition for these type of objects \citep{2013ApJ...775...10V, 2012A&A...539A..28G,2011ApJ...733....2N}. A more extended discussion concerning the different type of planets that could be formed when the enrichment of the envelope is considered, is presented in Sect.\ref{gasAccretion}.

\bigskip

\subsection{Dependence on the boundary conditions and semimajor axis}\label{test_CB}
All the results shown before were for a planet whose boundary conditions are those stated in Table\ref{tabCB}. In order to check the range in boundary conditions for which condensation can lead to a significant reduction of the critical mass, we perform the following test. We set the outer temperature and pressure of the planet for a given position in the protoplanetary disc making use of the following simple disc model \citep[see, namely,][]{2010apf..book.....A}: 

\begin{equation}\label{Tout}
 T_{\text{out}} = T_0 \bigg(\frac{a_0}{a}\bigg)^{1/2}
\end{equation}
\begin{equation}\label{Pout}
 P_{\text{out}} = P_0 \bigg(\frac{a_0}{a}\bigg)^{3}
\end{equation}
where $a_0$, $T_0$, $P_0$ are the standard boundary conditions used in all previous results, i.e, those given in Table\ref{tabCB}. In this section, we will refer to the simulations that use boundary conditions at $a = a_0$ as the ``reference case''.

\begin{figure}
 \includegraphics[width=\columnwidth]{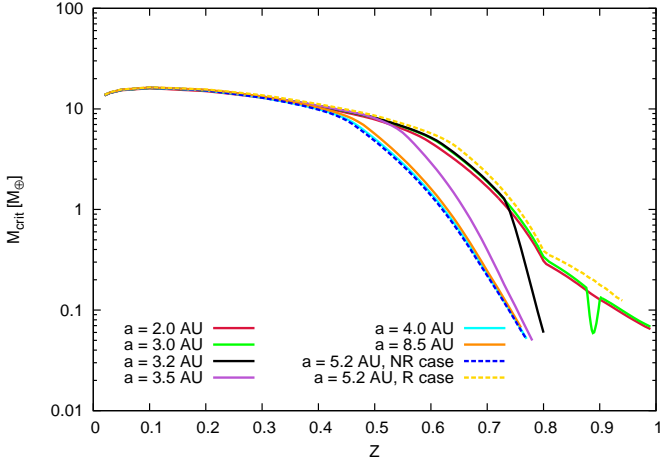}
 \caption{Critical core mass as a function of metallicity for the specified semimajor axis and corresponding boundary conditions given by Eqs. \ref{Tout} and \ref{Pout}. Dashed lines correspond to the reference cases (boundary conditions at $a=5.2$ AU) plotted in Fig. \ref{Mcrit_Z}: the yellow curve is the``restricted'' case (i.e., without condensation of water), and the blue one the ``non-restricted'' case, (where water condenses). For all the new cases (solid lines) water is allowed to condense. The only case where water does not condense for any value of $Z$ is for $a= 2$ AU.} 
 \label{Mcrit_BC}
\end{figure}

The results for the critical core mass as a function of metallicity for different semimajor axis (with the corresponding outer temperature and pressure given by Eqs.\ref{Tout}, \ref{Pout}) are shown in Fig.\ref{Mcrit_BC}. Results for 2 AU $\leq a \leq$ 8.5 AU are shown in solid lines, superposed with the reference cases in dashed lines (see caption). The qualitative similarity of the different curves with the reference case where water condenses (blue curve) or with the one where water is not allowed to condense (yellow curve) provides a hint on whether water condensation is taking place or not for the different semimajor axis considered. We note that for $a$= 2 AU, $T_{\text{out}}$ is always too high for water condensation to take place. For all the other curves condensation of water takes place at some point. The start of the condensation corresponds to the change in the slope of the curves. A particular case takes place for $a$ = 3.0 AU, where a valley in $M_{\text{crit}}$ occurs for $0.865 <Z< 0.905$ (see Fig.\ref{Mcrit_BC}). The drop in $M_{\text{crit}}$ corresponds to the fact that condensation starts to happen at $Z$ = 0.865. The same happens for $a$ = 3.2 AU at $Z$ = 0.75 and for $a$ = 3.5 AU at $Z$ = 0.55. The surprising thing about the 3 AU case is that for even higher metallicities, water condensation ceases. The reason for this is as follows: when the minimum in the ``valley'' of $M_{\text{crit}}$ is reached, the metallicity has increased so much that there is no more free H$_2$ at the top of the envelope. That is, H$_2$ starts to be the limiting reactant, so increasing Z (that is, adding more C and O) does not lead to the formation of more water. Instead, more CO$_2$  starts to be formed in this region of the valley (where d$M_{\text{crit}}$/dZ > 0). At $Z$ = 0.905, the amount of water formed is low enough to make the vapor pressure be lower than the saturation vapor pressure for all the \textit{P,T} values of the envelope. Thereby, for $Z\geq 0.905$, the profiles follow again the ``dry'' adiabat, no water condensation takes place anymore.

\begin{figure}
 \includegraphics[width=\columnwidth]{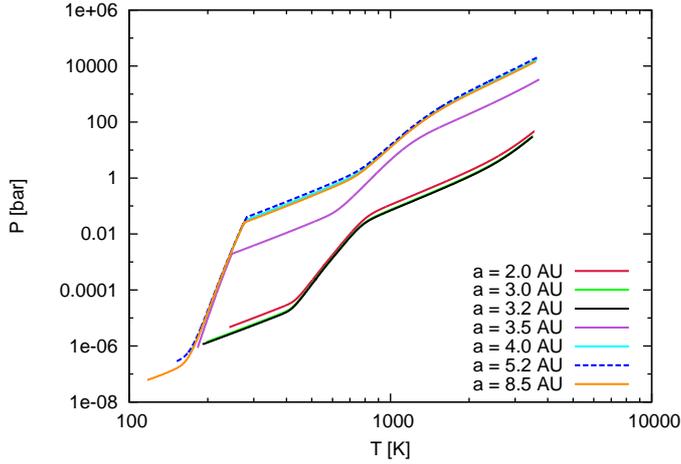}
 \caption{Pressure-temperature profiles for different position in the disc for Z=0.7 and $M_{\text{P}}=0.29 M_{\oplus}$. $T_{\text{out}}$ and $P_{\text{out}}$ correspond to those given by Eqs.\ref{Tout}, \ref{Pout} for the semimajor axis specified in the labels.} 
 \label{BC_all_effects_together}
\end{figure}

\begin{figure}  
\centering
\begin{tabular}{c}
\includegraphics[width=\columnwidth]{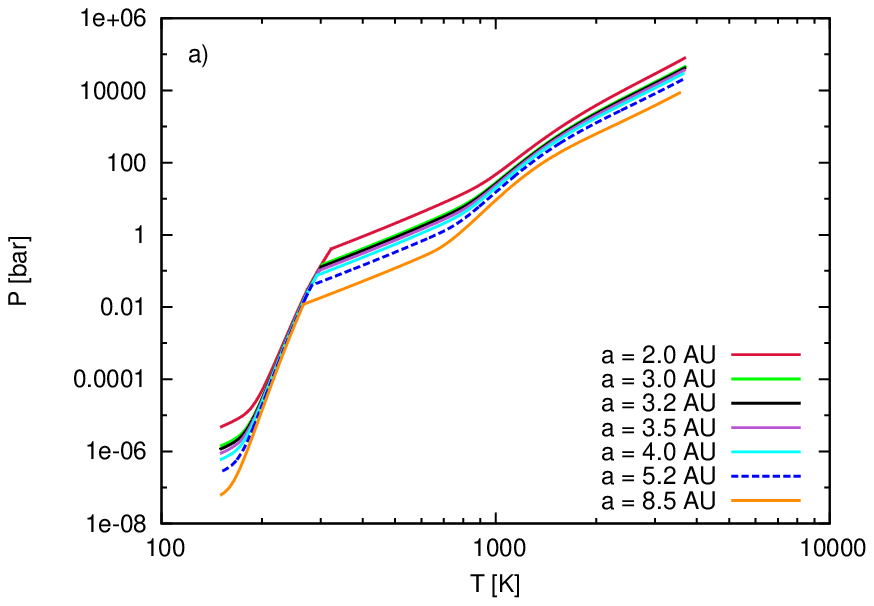}\\
\includegraphics[width=\columnwidth]{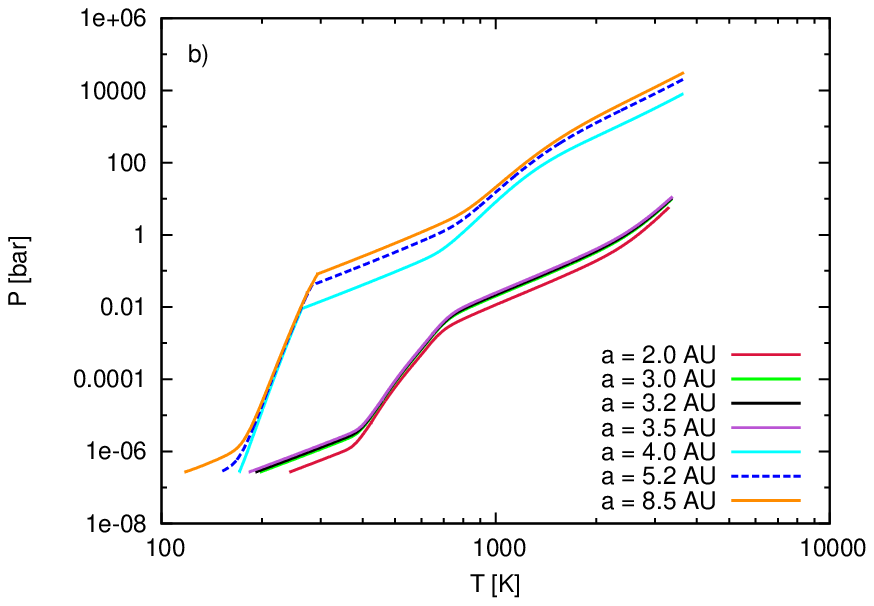}
\end{tabular}
\caption{Pressure-temperature profiles for different position in the disc for Z=0.7 and $M_{\text{P}}=0.29 M_{\oplus}$. a) $P_{\text{out}}$ given by Eq.\ref{Pout} for the labeled semimajor axis, but $T_{\text{out}}$= 150 K (reference case, see Table \ref{tabCB}) for all the profiles . b) $T_{\text{out}}$ given by Eq.\ref{Tout} for the labeled semimajor axis, but $P_{\text{out}}$ corresponds to the reference case for all the profiles.}
\label{profiles_BC}
\end{figure}

Despite this particular case of $a$ = 3 AU, the general trend observed in Fig.\ref{Mcrit_BC} is that H$_2$O condensation occurs more readily for planets located further out in the disc. However, while the trend is clear, the exact behavior is more difficult to predict. For instance, the behavior of $M_{\text{crit}}$ changes considerably for $3\leq a \leq 4$ AU, but remains very similar for the cases of $a$ = 4, 5.2 and 8.5 AU. To have a better insight on the difference in $M_{\text{crit}}$ among these cases, we analyse profiles for $Z = 0.7$ and $M_{\text{P}} = 0.29 M_{\oplus}$ for all the different cases shown in Fig.\ref{Mcrit_BC}. The results of changing $a$ and the corresponding $P_{\text{out}}$ and $T_{\text{out}}$ (as dictated by Eqs.\ref{Tout}, \ref{Pout}) are shown in Fig.\ref{BC_all_effects_together} \footnote{If migration were taken into account, this change in the boundary conditions would be the way to compute the internal structure of the planets in the case that immediate adaptation to the disc temperature and pressure values were assumed.} 
As expected from the behavior of $M_{\text{crit}}$ at $Z = 0.7$ in Fig.\ref{Mcrit_BC}, the cases of $a \leq 3.2$ AU follow the ``dry'' adiabat, whereas the cases of $a \geq$ 3.5 AU correspond to the ``moist'' adiabat. Since there are several factors that could be playing a role in shaping these profiles, we study the effect of changing just one boundary condition and fixing all the others with the standard values (i.e, those given in Table\ref{tabCB}) in Figs.\ref{profiles_BC} a,b. 

Fig.\ref{profiles_BC}a shows the effect of changing the outer pressure, $P_{\text{out}}$. That is, the $P_{\text{out}}$ used to solve the internal structure equations is the one given by Eq.\ref{Pout}, but all the other boundary conditions correspond to the standard case. From this figure we infer that the critical core mass is reduced when $P_{\text{out}}$ increases, since the pressure in the envelope is always higher for higher $P_{\text{out}}$. This fact can be understood as follows: from Eqs. \ref{implicit}, \ref{dTdPs} and \ref{cp}; it can be inferred that $(\frac{\partial P}{\partial S})_T < 0$. Hence, at the outer boundary of the planet, the entropy decreases for larger $P_{\text{out}}$. For these profiles of Z=0.7, the envelopes are fully convective, so the entropy of the envelope is fixed by $P_{\text{out}}$, $T_{\text{out}}$. On the other hand, the temperature at the surface of the core is the same for all the cases because they correspond to the same planetary mass\footnote{Property 
that can be inferred for stellar structure making use of the Virial Theorem, see, for instance, \citet{2013sse..book.....K}. For bodies with a core this is not strictly true, but the changes in central temperature are negligible.}. Therefore, the pressure at the surface of the core also has to be larger for larger $P_{\text{out}}$ \citep[see][for a discussion on the sensitivity of the critical core mass with the outer boundary conditions for fully convective envelopes]{1993Icar..106..323W, 2001ApJ...553..999I}.

In Fig.\ref{profiles_BC}b, the only parameter we change is $T_{\text{out}}$. The first remarkable observation is that if $T_{\text{out}}$ is larger than the temperature given by the Clausius-Clapeyron relation at $P_0$, no condensation of water occurs. In our particular disc toy model this happens for $a \leq 3.5$ AU. Then, the change in $T_{\text{out}}$ is what gives raise to the two types of profiles observed in the general case (Fig.\ref{BC_all_effects_together}, where all boundary conditions are modified consistently with Eqs.\ref{Tout}, \ref{Pout}). 
The general trend of changing $T_{\text{out}}$ is that $M_{\text{crit}}$ diminishes when $T_{\text{out}}$ diminishes. This can be explained drawing similar arguments than with Fig.\ref{profiles_BC}a. Due to Eq.\ref{cp}, we know that  $(\frac{\partial S}{\partial T})_P > 0$, and hence, at the outer boundary of the planet the profiles with lowest $T_{\text{out}}$ are the ones with lowest entropy. Since the temperature at the surface of the core is the same for all the profiles, the profiles with lowest entropy will have the largest central pressure (because $(\frac{\partial P}{\partial S})_T$ is always negative). Consequently, regarding $T_{\text{out}}$, the further the planet is from the star, the smaller will be $M_{\text{crit}}$.

We have found, therefore, that the change of $P_{\text{out}}$ and $T_{\text{out}}$ provoke opposite effects in the change of the critical core mass as we move further out in the disc. This is what makes Fig.\ref{Mcrit_BC} and Fig.\ref{BC_all_effects_together} not so easy to interpret. Looking back at the profiles of Fig.\ref{BC_all_effects_together}, we see that the case of $a$ =3.5 AU is the only one in which the profile is moist due to the increase in $P_{\text{out}}$ (when comparing to the reference case), which place the pressure-temperature curve in the left side of the Clausius-Clapeyron curve. For the cases of $a$ =2, 3 and 3.2 AU, the increase in $P_{\text{out}}$ that actually occurs with respect to Fig.\ref{profiles_BC}b is not enough to switch the profiles from dry to moist. Hence, in these cases, what makes these profiles follow the ``dry'' adiabat (i.e, water does not condense) is the increase in $T_{\text{out}}$ with respect to the reference case.

Concerning the cases of $a$ = 4, 5.2 and 8.5 AU, since ($T_{\text{out}}$,$P_{\text{out}}$) reside always in the region where H$_2$O is in solid phase (or in the solid-vapor transition), the profiles follow the moist adiabat. Due to the opposite effects on $M_{\text{crit}}$ that arise when $P_{\text{out}}$ and $T_{\text{out}}$ decreases as the planet is further from the star, the profiles in these cases are very similar.

We can conclude then that the decrease in $M_{\text{crit}}$ with increasing metallicity due to the condensation of water in the envelope of protoplanets is robust with respect to changing boundary conditions. The exact location in the disc where H$_2$O starts to condense will depend on the disc model and stellar spectral type, but for distances larger than this, the value of $M_{\text{crit}}$ will be practically independent of the location of the planet in the disc, as shows Fig.\ref{Mcrit_BC} for a $\geq$ 4 AU.

\section{Discussion}\label{disc}

\subsection{Validity of the uniform metallicity assumption} \label{uniformZ}
Although the disruption of planetesimals is more likely to occur in the densest regions close to the surface of the core, the final fate of the ``metals'' is not a priori clear. The heavy elements could sink slowly to the core, or get well mixed in the H-He envelope, depending upon the strength of convection and the solubility of the material in the H-He envelope.
In Sect. \ref{results} we assumed that the metallicity was uniform throughout the envelope, and this may not be always the case. Nevertheless, if the envelope is fully convective, this assumption is a reasonable one. This was the case for all envelopes with $Z \ge 0.45$. However, it could also be that the envelopes were fully convective because a uniform Z was assumed. It is important to know if this is the case or not, because if the enrichment does not reach the lowest temperature region of the envelope, no water will be present for temperatures lower than its condensation temperature, and then the condensation of water will not take place (at least for the boundary conditions considered in this work). 

To evaluate how well mixed the envelope will be, we performed the following test. We assumed that the enrichment of material occurs at temperatures higher than the condensation temperature of water. Hence, the outermost part of the atmosphere ($T<T_{\text{cond}}$) is assumed not to be enriched and have a solar composition, whereas the innermost part ($T>T_{\text{cond}}$) is assumed to have a uniform Z (higher than solar).\footnote{This scenario could take place if the water droplets formed where $T<T_{\text{cond}}$ precipitated, which is the reason why we chose $T_{\text{cond}}$ as the boundary between the polluted and non polluted layers.} If with this two layered model atmosphere the envelopes are still fully convective, then a global mixing of material will occur and therefore the envelopes will acquire, eventually, a uniform composition.

Since this initial set-up implies a considerable jump in mean molecular weight at $T = T_{\text{cond}}$ (where the boundary between the solar composition and the high-Z region occurs), the Ledoux criterion for stability against convection should be used \citep[][chap. 6]{2013sse..book.....K}, instead of the Schwarschild criterion. Therefore, for our calculation the radiative zones are those for which:
\begin{equation}
 \nabla_{\text{rad}} < \nabla_{\text{ad}} + \nabla_{\mu}
\end{equation}
where $\nabla_{\mu}=\frac{\varphi}{\delta} \bigg(\frac{\partial \text{ln} \mu}{\partial \text{ln} P}\bigg)_{env}$, $\varphi= \bigg(\frac{\partial \text{ln} \rho}{\partial \text{ln} \mu}\bigg)_{P,T} $ and $\delta = -\bigg(\frac{\partial \text{ln} \rho}{\partial \text{ln} T}\bigg)_{P, \mu}$. For an ideal gas $\varphi= \delta = 1$.

Due to the finite jump of $\mu$ at $T = T_{\text{cond}}$, $\nabla_{\mu}$ diverges at this temperature. Hence, to compute the composition gradient, we assumed the same simple prescription used in stellar evolution models to treat the problem of overshooting \citep{1999A&A...344..551A}, i.e, we assumed that the change of $\mu$ is effective within a distance $0.2 H$ of the radius at which $T=T_{\text{cond}}$ (H is the scale height of the atmosphere at $T=T_{\text{cond}}$), and we take $\nabla_{\mu}$ constant in this interval. 

For the opacity of the solar composition layer, we use, for the sake of consistency, the same opacities that we used before, but with a $Z=Z_{\odot}$ for the Ferguson gas opacities. It is important to remark that for the range of temperatures of this (predominatly) H-He region, the Ferguson opacities are extrapolated, and therefore, its values are probably not very accurate. We will discuss the effect of varying this opacity in the next section. 

The results of the test show that the critical envelopes of the standard, non-restricted case remain fully convective when a layer of solar composition is considered for $T<T_{\text{cond}}$, provided that $Z \ge 0.6$. This means that for these high enriched cases, the assumption of uniform Z was a good one, and hence that condensation of water is likely to take place, at least for the opacities considered here. 

Finally, whatever the fate of the water droplets (this depending on the dynamics of the atmosphere and the miscibility of the forming droplets), if the envelope remains convective up to the layer where H$_2$O can condense, it is likely that the continuous saturated state of  H$_2$O can sustain a moist-adiabatic pressure-temperature profile.\footnote{Actually, the condition of a fully convective envelope to have perfect mixing is overly severe. In reality, vertical mixing tends to take place in regions well above the convective-radiative boundary. On Earth, namely, the convective region extends for 10 km over the planet surface, but vertical, efficient mixing of material occurs up to 80 km height \citep{1995papc.book.....G}.} This will make it possible for the critical core mass to reduce considerably, as shows Fig.\ref{Mcrit_Z}. Therefore, for high enriched envelopes \citep[which should not be so rare during the formation scenario, as shown by][]{2013ApJ...775...80F}, the effect of condensation of water can really play a role.

\subsection{Role of microphysics}

\subsubsection{Ideal gas assumption}\label{validity_CEA}
Since in this work an ideal gas equation of state was assumed for the structure calculation, it is relevant to have an estimate of the error introduced by neglecting the degeneracy of the gas at high pressure. For this reason, a comparative test for H-He envelopes implementing CEA and SCVH EOS was performed.

\begin{figure}
\begin{center}
\includegraphics[width=\columnwidth]{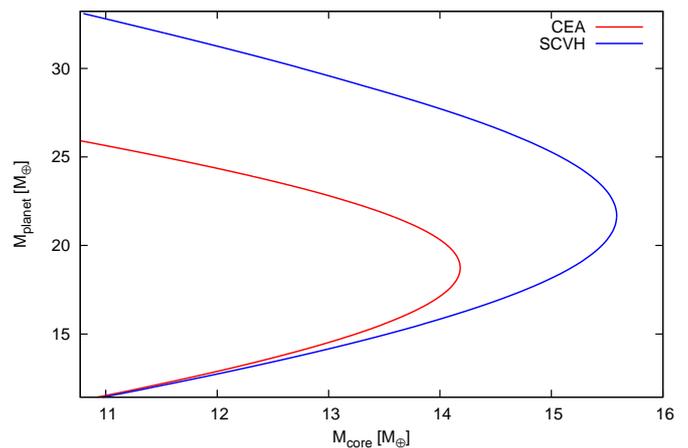}
   \caption{Planetary mass as a function of the core mass given by solving the internal structure equations for pure H-He envelopes. The red curve uses CEA for the EOS and the blue, SCVH. Boundary conditions are those from Table \ref{tabCB}. $\Delta M_{\text{crit}} \approx$ 1.4 $\text{M}_{\oplus}$ }
    \label{Mcrit_HHe}
\end{center}
\end{figure}

The results of the planetary mass as a function of the core mass (obtained by solving the internal structure equations) are shown in Fig. \ref{Mcrit_HHe}. From this figure, it can be appreciated that differences in planetary mass for a given core mass start to be noticeable when reaching the critical core mass. The critical core masses are 14.18 $\text{M}_{\oplus}$ and 15.58 $\text{M}_{\oplus}$, for CEA and SCVH cases, respectively; which implies a relative error of $\sim 10 \%$ in the computation of the critical core mass. Even performing the same test with other boundary conditions or opacities, we find that the error is never larger than $\sim 12 \%$ \footnote{Note, however, that due to the non-linearity of the planet formation process, this difference in critical core mass can significantly influence the final characteristics of the planet.}

\begin{figure}
\begin{center}
\includegraphics[width=\columnwidth]{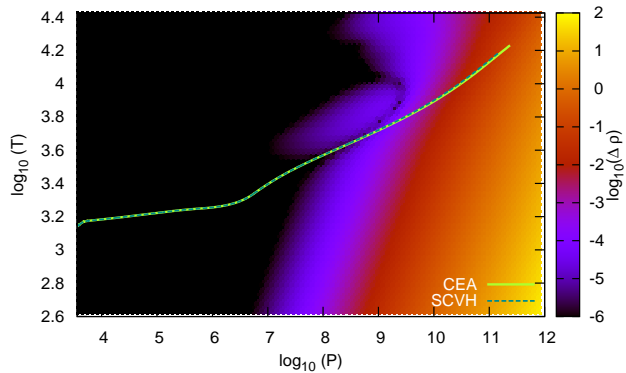} 
   \caption{Map of the density difference between CEA and SCVH EOS as a function of temperature and pressure, for pure H-He composition. Superposed are shown the profiles of the envelope for the two different EOS for $M_{\text{P}}$ = 18.7 $\text{M}_{\oplus}$ (planetary mass for which $M_{\text{core}}=M_{\text{crit}}$ when CEA is used for the EOS).
   All units in cgs.}
    \label{map}
\end{center}
\end{figure}

To have a better insight of the effect of degeneracy, we show, in Fig.\ref{map}, a map of the difference in density obtained using CEA and SCVH EOS, as a function of (log P, log T). Overlapped are the profiles of the envelope obtained for $M_{P} = M_{\text{P}}(M_{crit, CEA})$ for CEA and SCVH cases (same conditions as in Fig.\ref{Mcrit_HHe} were assumed). Although the difference between CEA and SCVH profiles is not easily distinguishable, it can be noticed that the CEA profile overestimates the central temperature and pressure, leading, as a consequence, to an also overestimated central density. The difference in central density between these CEA and SCVH profiles is of 0.1 g/cm$^3$. The higher density obtained using CEA translates into a higher envelope mass, and hence, in a smaller critical core mass, as shows Fig.\ref{Mcrit_HHe}. 
We have shown that the change in critical core mass induced by considering the enrichment of the envelope (and in particular, the enrichment by condensable species) is orders of magnitude larger than the changes produced by including non-ideal effects. This justifies our assumption in assessing the effect of condensation on the determination of the critical mass.

\bigskip

\bigskip

\subsubsection{Remarks on the opacity}\label{remarks_opac}
\begin{figure}
 \includegraphics[width=\columnwidth]{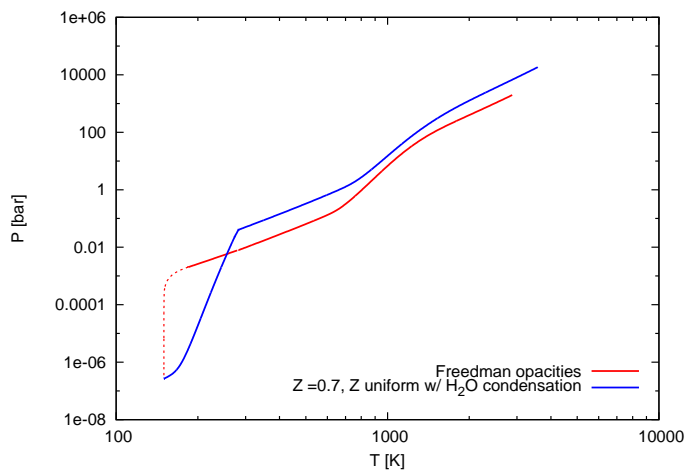}
 \caption{red line: profile for two-layered atmosphere, but using Freedman opacities in the H-He region instead of the Ferguson opacities used in all other results of the paper. Blue line: Z=0.7 (and uniform in the envelope) where water is allowed to condensed. The resemblance of both profiles, compared to the ``restricted'' profile of Fig.\ref{Z07_Pvapor}, suggests that for the red curve the corresponding critical core mass will also be very low. Solid lines stand for convective regions and dotted for radiative ones.}
 \label{FREED}
\end{figure}

Concerning the opacity, it should be kept in mind that the Ferguson opacities for the gas are extrapolated in the coldest layers of the envelopes computed in this work. In principle this does not affect the profile of envelopes with high and uniform metallicity, since they are fully convective; but it could affect the structure for the test described in Sect.\ref{uniformZ}. For instance, if instead of using the extrapolated Ferguson opacities for the H-He region, we used \citet{2008ApJS..174..504F} tables, the envelope would become radiative in the H-He region. This is because the Freedman opacity values are lower for the cold regions than the ones extrapolated from Ferguson, and hence, the adiabatic gradient becomes larger than the radiative one. In this case, then, the mixture of the material would probably not occur in the whole envelope. However, because of the use of reduced opacities in the outermost part of the atmosphere, the critical core mass would still remain very low \citep{2000ApJ...537.1013I,
Hubi05,2010ApJ...714.1343H}, as suggested by Fig.\ref{FREED}. It is interesting to remark that even if condensation of water does not take place, the critical core masses shown in Fig.\ref{Mcrit_Z} will not be significantly affected.

\subsection{Chemical equilibrium assumption}
There are basically two processes that can preclude chemical equilibrium in planetary atmospheres: UV-photochemistry in close-in exoplanets, and vertical mixing in cold atmospheres of giant planets \citep[][and references therein]{2014RSPTA.37230073M}. It is well known that in Jupiter, the measured CO abundance at the top of the atmosphere is much higher than expected from chemical equilibrium calculations \citep{Lod10}. The main reaction producing CO in planetary atmospheres is $ \text{H}_2\text{O} + \text{CH}_4 \rightleftarrows 3 \text{H}_2 + \text{CO} $. The timescale of this reaction is indeed so long for the temperature and pressure at the top of Jupiter's atmosphere, that CO is no longer destroyed by this reaction \citep{2010eapp.book.....S}. The amount of CO measured at the top of the atmosphere corresponds to that at T$\sim$1000 K, where chemical equilibrium still holds \citep{Lod10}. The CO produced at this level is carried outwards by vertical mixing. Hence, the CO is froze-in, or ``quenched'' at T $\sim$1000 K (the so-called quench level).

The aforementioned reaction affects as well the amount of water at the top of the atmosphere, issue of major importance for our results, since the effect of water condensation on the critical mass depends on how much water is present in the cold layers of the envelope.

The real amount of water will depend on several mechanisms operating in a planet formation scenario: the amount present in the incoming planetesimals, the depth at which they are completely destroyed, the strength of convection and the extent of the convective cells, etc. To simplify this complex scenario, let us imagine the following picture. We ignore where planetesimals deposit their mass. We assume large scale-convection. Hence, the volatiles provided by planetesimals (whatever their molecule/atomic state) will also be transported to the top of the core, where temperature is high enough to dissociate any water molecule. Then, due to convection, the bubbles reach the quench level and the abundance of water will remain fixed in that amount up to the top of the atmosphere. Therefore, in order to test in an approximate way the effect of disequilibrium, we assume in the test of this section that the amount of water is the one in equilibrium at a quenching temperature. In reality, the quench level depends on the species present in the atmosphere and their concentrations. Given the similarity of the P-T profiles in our planets and in Jupiter around T=1000 K, we take $T_{\text{quench}}$=1000 K.

In this section we focus on the abundance of water, which is the key factor in our results. For this, we consider a fictitious atmosphere that is made only of water and noble gases. Since no reaction occurs between water and noble gases, this scenario provides a good opportunity to study in an approximate way the effect of water condensation in a system where all the species remain quenched until the temperature is high enough for water to dissociate (i.e, for T$\lesssim$2000 K). We mimic, in this way, what would happen in a disequilibrium scenario.  

\begin{figure}
 \includegraphics[width=\columnwidth]{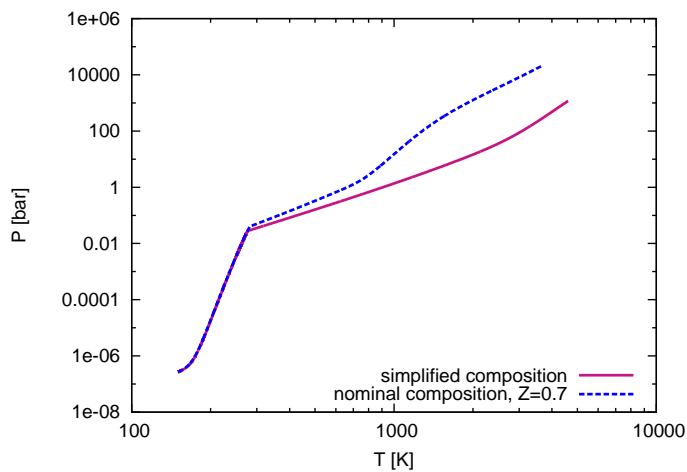} 
 \caption{Comparison of pressure-temperature profile between the simplified envelope composition and the nominal composition for Z=0.7. In both cases water is allowed to condense. The same mole fraction of water at the top of the atmosphere is assumed for both profiles. Therefore, the departure from the moist adiabat occurs at the same (P,T) in both profiles.}
 \label{PT_water_noble}
\end{figure}

In Fig.\ref{PT_water_noble} we present temperature-pressure profiles for the ``nominal'' composition case (the envelope composition assumed so far in this work, i.e, H, He, C and O) and the new ``simplified'' case, where no chemical reactions take place since the envelope is made of water and noble gases\footnote{The noble gases considered  for the ``simplified composition'' are He and Ne and the mole fractions are chosen to have approximately the same mean molecular weight at the top of the atmosphere than in the ``nominal'' composition case.}. Both cases share the same mole fraction of water at the top of the atmosphere. As was explained in Sect.\ref{results}, this guarantees that the departure from the moist-adiabatic profile occurs at approximately the same (P,T). Notice, however, that the profiles follow different tracks after the departure from the moist adiabat. This is due to the lack of chemical reactions in the simplified case. In the nominal case, chemical reactions take place at T$\gtrsim$ 500 K. 
This lowers the adiabatic gradient and causes the bump of the blue curve in that range of temperature. Hence, computing the critical core mass with the simplified envelope composition provides us not only a way of simulating a quenched (or disequilibrium) scenario, but also allows us to disentangle the effect of reduction of critical core mass due to chemical reactions and water condensation.

\begin{figure}
 \includegraphics[width=\columnwidth]{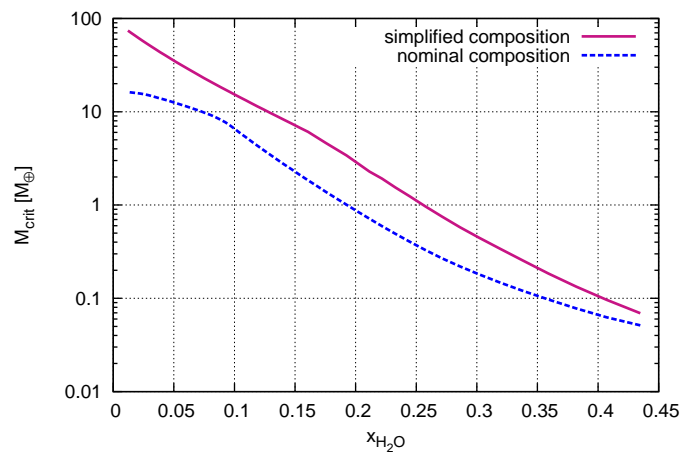} 
 \caption{Critical core mass as a function of the mole fraction of water in the simplified case (purple) and nominal case (blue). Water condensation takes place in both cases.}
 \label{Mcrit_xtop}
\end{figure}

\begin{figure}
 \includegraphics[width=\columnwidth]{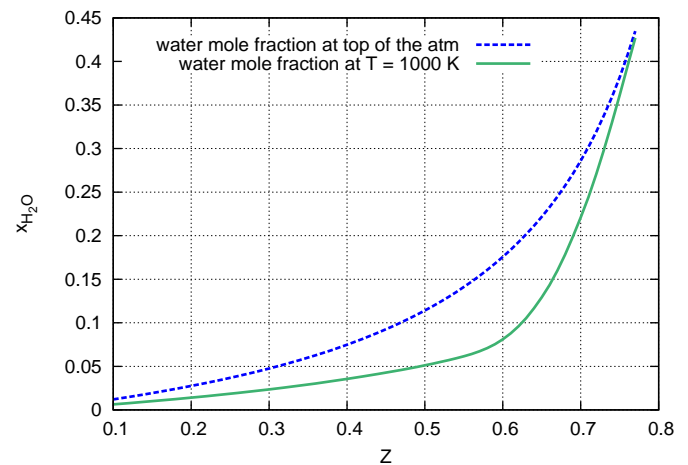} 
 \caption{Nominal composition case. Relation between the metallicity and the mole fraction of water at the top of the atmosphere (blue) and at T=1000 K (green).}
 \label{Z_xtop}
\end{figure}

The effect of water condensation is determined by the point in the temperature-pressure diagram at which the vapor pressure stops being saturated (and this depends just on the mole fraction of vapor). Thus, concerning the effect of water condensation on the critical core mass, the two profiles shown in Fig.\ref{PT_water_noble} are equivalent. We can therefore compute the critical core mass for the simplified case in two steps:

\begin{enumerate}
 \item We compute the critical core mass as a function of the mole fraction of water ($x_{\text{H}_2\text{O}}$) for the simplified composition (Fig. \ref{Mcrit_xtop}). 
 \item We compute the mole fraction of water as a function of Z for the nominal composition (Fig.\ref{Z_xtop}, purple curve).
\end{enumerate}

Hence, by combining Figs. \ref{Mcrit_xtop} and \ref{Z_xtop} (purple curve) we can express the critical core mass of the simplified composition as a function of an ``equivalent'' Z, for which ``equivalent'' means having the same mole fraction of water as the one on the top of the atmosphere of the nominal composition. For instance, for a mole fraction of water of 0.29 in the simplified case, $M_{\text{crit}} \approx 0.5 \text{M}_{\oplus}$ (Fig.\ref{Mcrit_xtop}). For the nominal composition, this mole fraction of water at the top of the atmosphere corresponds to a metallicity of Z=0.7. Thus, the simplified composition with a mole fraction of water of 0.29 has an``equivalent metallicity'' of 0.7.

\begin{figure}
 \includegraphics[width=\columnwidth]{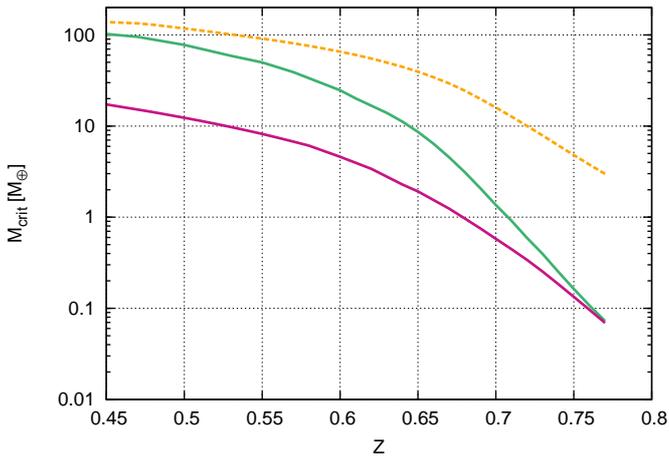} 
 \caption{Critical core mass as a function of the equivalent metallicity for the simplified composition cases. Purple solid line: case where water condenses and the mole fraction of water is the same as the one at the top of the atmosphere of the nominal composition case (i.e, the one given by the purple line of Fig.\ref{Z_xtop}). Green-solid: water condenses but the mole fraction of water corresponds to that at T= 1000 K of the nominal composition (green line of Fig.\ref{Z_xtop}). Orange-dashed: same as green curve but in this case water is NOT allowed to condense.}
 \label{Mcrit_Z_simplified}
\end{figure}

Nevertheless, as was explained before, the fact of having $x_{\text{H}_2\text{O}}$=0.29 at the top of the atmosphere in the nominal case for Z=0.7 is a result of assuming chemical equilibrium in that layer, which is probably not true. Thus, a more realistic amount of water at the top of the atmosphere should be the one at T=1000 K (namely, $x_{\text{H}_2\text{O}}$=0.22 for Z=0.7, see green curve of Fig. \ref{Z_xtop}), where chemical equilibrium is a reliable assumption. We therefore compute again the critical core mass for the simplified composition, but assuming that the mole fraction of water is the one of the nominal composition at T=1000 K. The critical core mass as a function of the equivalent metallicity for the simplified compositional cases (with and without water condensation) are shown in Fig.\ref{Mcrit_Z_simplified}. \footnote{For simplification, fully convective envelopes were assumed. Since this is the case for the nominal composition for Z $\geq$ 0.45, we just show results for this range of 
metallicities, so comparison with Fig. \ref{Mcrit_Z} is suitable.} 

From Fig.\ref{Mcrit_Z_simplified} we can draw the following conclusions. In a chemical disequilibrium scenario (like the one we mimic with our simplified composition), considering or not the condensation of water (green and orange curves of Fig.\ref{Mcrit_Z_simplified}, respectively), brings a difference in critical core mass of still an order of magnitude for large Z, the same that was found considering chemical equilibrium (Fig.\ref{Mcrit_Z}). Of course, the absolute values of $M_{\text{crit}}$ are larger in the disequilibrium scenario, because the reduction of adiabatic gradient we had for T$\gtrsim$ 500 K (Fig. \ref{ad_grad}) due to chemical reactions does not take place in these water/noble gas envelopes. With this simplified composition we are excluding all chemical reactions (except dissociation of H$_2$O). This is not justified for T$\gtrsim$ 1000 K, and also, in reality, the presence of dust grains may increase the rate of chemical reactions \citep{Fegley94} throughout the 
envelope. Therefore, the absolute $M_{\text{crit}}$ values of Fig.\ref{Mcrit_Z_simplified} are upper limits when translating these into the nominal-composition out-of-chemical equilibrium case.

Note finally that we assumed in this test that all water condenses when the pressure and temperature are suited for it to occur. This is true in an equilibrium scenario. It is well known within the cloud formation community that in the presence of only one condensible species, the creation of the first condensates -nucleation- might require a high level of supersaturation \citep[see, e.g., the reviews of][]{HellingRev, MarleyRev}. However, if the atmosphere contains preexisting nucleating seeds, like dust grains, condensations can already occur for saturation ratios close to one. This heterogeneous nucleation is what explains cloud formation on Earth. In the atmospheres we are referring in this work, dust grains are much more abundant than on Earth due to the constant replenishment by the ablation of planetesimals. Therefore, water can effectively nucleate and condense in the atmospheres of these growing planets. 

\subsection{Condensation of other species}\label{ammonia}
Apart from water, other compounds could condense in the pressure-temperature regimes considered in this work. In particular, it is known that the atmosphere of Jupiter possesses NH$_3$ clouds. Unfortunately, while CEA does handle NH$_3$ in gaseous form, it does not take into account its condensation. However, we can at least test whether the abundance of NH$_3$ in the atmosphere would be sufficient for its condensation to matter. For this, we consider an atmosphere in which we add nitrogen to the already included H, He, C an O. We keep the same solar C/O as used throughout, and assume also a solar ratio of N/O. Using CEA, we find that in the coldest layers of the envelope, the mass fraction of NH$_3$ formed is just $\sim 5\%$ . Therefore, even if all the ammonia formed could condense, this would be a minor contribution compared to the amount of condensed water. Even more important, if condensation of other species took place, the effect of reduction of critical core mass would be even larger, since 
the effect of release of latent heat explained in Sect.\ref{results} is not unique to water, but qualitatively general to all vapor-solid (or vapor-liquid) phase transitions.

\subsection{Gas accretion timescale}\label{gasAccretion}
The growth of an enriched, critical protoplanet is not easy to infer. A simple estimation of the gas accretion timescale can be obtained by computing the Kelvin-Helmholtz timescale, which corresponds, approximately to \citep{2000ApJ...537.1013I} :
\begin{equation}
\tau_{KH} \sim \frac{G M_{\text{core}} M_{\text{env}}}{R_{\text{core}} L} 
\end{equation}
Using Eq.\ref{Mdotcore}, this can be simplified to:
\begin{equation}
\tau_{KH} \sim \frac{M_{\text{env}}}{\dot{M}_{\text{core}}} 
\end{equation}

Evaluating this expression when the structures become critical, and using $\dot{M}_{\text{core}} = 10^{-6} M_{\oplus}/\text{yr}$ (which is the approximated planetesimal accretion rate for the cases presented in this paper, as well as a standard value in the planet formation literature, see e.g., \citet{1986Icar...67..391B, 1996Icar..124...62P, 2000ApJ...537.1013I}), we find, for Z=0.7, that $\tau_{KH} \sim 7 \times 10^4 \text{yr}$ when water condensation takes place, and  $\tau_{KH} \sim 10^6 \text{yr}$ when water does not condense. In principle, these timescales would indicate that the condensation of water significantly speeds up the accretion of gas. However, two things have to be kept in mind:
\begin{enumerate}
 \item Accretion rate of gas depends on both the self-gravity of the envelope, and its ability to cool (and thereby, to contract). The latter depends on the microphysics of the envelope. Solving the full planet-growth problem for enriched envelopes would imply, in part, considering self-consistent grain growth calculation for the opacity \citep{2014ApJ...789L..18O, 2014arXiv1406.4127M}, because this affects the planet cooling. This, coupled to a code that solves the structure with a varying luminosity with time, will be implemented in a future work.
 
 \item Even if the gas accretion timescale is shorter for the case where water condensation occurs, gas accretion would lead to a decrease of the envelope metallicity (i.e. dilution), and therefore to the increase of the critical core mass.
\end{enumerate}

\begin{figure}
\begin{center}
\includegraphics[width=\columnwidth]{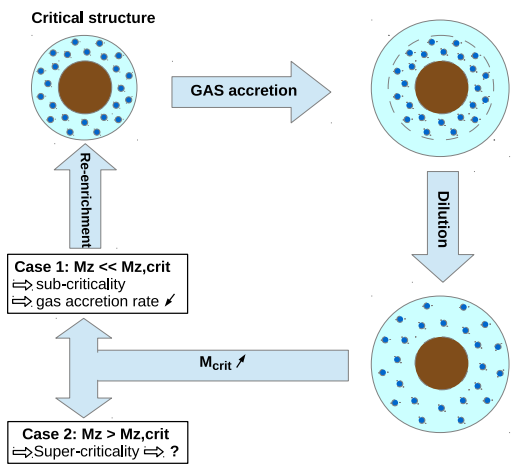}
   \caption{Sketch of the possible phases an enriched protoplanet could follow after reaching the critical mass. Although not illustrated (for not overloading the image), planetesimal accretion is present in all the stages and the material assumed to remain in the envelope. Despite the competition between gas and planetesimal accretion, the immediate result would be dilution, since at this point the accretion of gas is larger than that of solids, as explained in the main text. As a consequence of dilution, $M_{\rm crit}$ increases. After this point, depending on the accretion rate of planetesimals (which gives a new $M_Z$) versus the accretion rate of gas from the disc, the planet could become sub-critical or super-critical, as is discussed in Sect.\ref{gasAccretion}.}
    \label{sketch}
\end{center}
\end{figure}

Indeed, if a planet reaches the critical mass at a given $Z_{\rm crit}$, two effects will appear. First, as a result of solid (i.e., planetesimal) accretion, the metal content of the planet ($M_Z$) increases. Second, as a result of gas accretion, the mean $Z$ of the planet will change. According to \citet{2000ApJ...537.1013I}, the gas accretion rate at the critical point is similar, but slightly larger, than the planetesimal accretion rate. At the time when the critical core mass is reached, the accretion of solids and gas will lead to a decrease of the mean $Z$ in the envelope.
As a result, the critical core mass, as well as the total amount of heavy elements for the critical structure $M_{Z,\rm crit}$, increase. A key point, for the immediate future of the planet is therefore to compare the increase of $M_Z$ (resulting from the accretion of planetesimals) and the increase in the critical amount of heavy elements.\footnote{We are assuming in this discussion (as is illustrated in Fig. \ref{sketch}), that the mass of the core is fixed. Hence the heavy elements provided by planetesimals accretion remain in the envelope, making Z, and thereby $M_{Z,\rm crit}$, grow.}

Thus, two possible scenarios could take place after the critical core mass is reached (see sketch on Fig. \ref{sketch} for an illustration). In the first scenario (case 1), the increase of critical mass is more rapid than the increase of $M_Z$. In this case, the planet will become again subcritical, will not accrete anymore gas. Finally, as a result of planetesimal accretion, the metallicty may increase again. In the second scenario (case 2), the increase of critical mass is smaller than the increase of $M_Z$. In this case, the planet will remain super-critical, and might become a giant planet. 

\begin{figure}
\begin{center}
\includegraphics[width=\columnwidth]{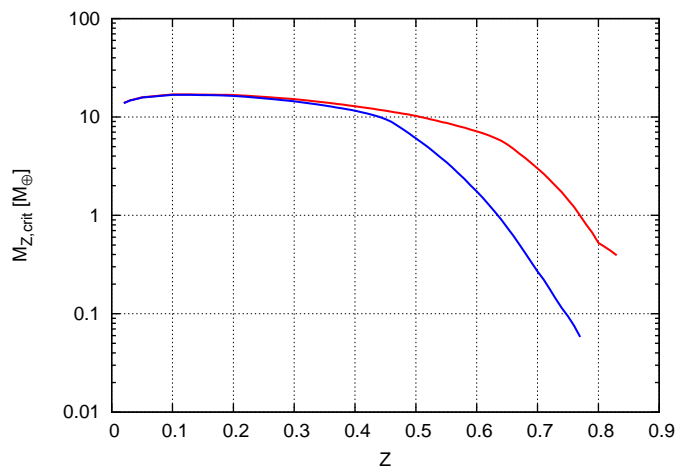}
   \caption{Critical mass of metals (mass content of metals in the planet when $M_{\text{core}}=M_{\text{crit}}$) as a function of metallicity. Red line: restricted case. Blue line: non-restricted case. Boundary conditions given by Table \ref{tabCB}. Note that despite the similarity with Fig.\ref{Mcrit_Z}, in this case the y-axis represents all the metal content of the planet at criticality, and not just that of the core. Depending on the slope of $M_{Z,\rm crit}(Z)$, for the same planetesimal accretion rate, the planet could become sub or super-critical. The steeper the slope (water condensation case), the more likely is the structure to become sub-critical, and thereby, to follow case 1 illustrated in the Fig. \ref{sketch}.}
    \label{Mzcrit_Z}
\end{center}
\end{figure}

The occurrence of case 1 or 2 depends therefore on the relative increase of $M_{Z,\rm crit}(Z)$ and $M_Z(Z)$. The steeper the slope of $M_{Z,\rm crit}(Z)$, the larger the increase of  $M_{Z,\rm crit}$ for a given reduction of $Z$. Thus, the higher the probability to be in case 1. For instance, as shown in Fig.\ref{Mzcrit_Z}, when comparing the case where water condenses with the one where water does not condense, the former case has a steeper slope than the latter. Therefore, water condensation would lead probably to case 1, and hence to an increase of the heavy element content of the planet.  

In reality, which case takes place depends on the location of the forming planet, solid accretion rate, among other conditions.

\section{Conclusions}\label{conclusions}
The critical core mass for envelopes enriched in carbon and oxygen through the disruption of planetesimals was computed, corroborating the reduced values found by \citet{2011MNRAS.416.1419H}. It was found that large amounts of water are formed in the envelope of the enriched protoplanets, and that an important fraction of it can condense for boundary conditions suited for $a \gtrsim$ 4 AU. Because of the release of latent heat that takes place when H$_2$O condensation  occurs, the energy budget of the envelope is dramatically changed. This translates into a drastic pattern shift in the pressure-temperature envelope profile, which implies a remarkable increase in pressure for a given temperature of the envelope. Therefore, compared to the case where H$_2$O condensation is not included, the density of the envelope in the condensed case increases significantly, reducing the critical core to extremely low values, as shows Fig.\ref{Mcrit_Z}.

The reduction of critical core mass due to water condensation depends on the amount of water present in the outer layers of the envelope, which is affected by the assumption of chemical equilibrium. However, we have shown that for a disequilibrium scenario, the critical core mass is still significantly reduced.

This critical core mass is important in the context of the ‘standard’ models of planet formation because it sets the time required to form a gas giant. Indeed, as shown in many papers, the gas accretion rate for a critical mass of $\sim$ 10 ${\mathrm M}_{\oplus}$ is so large that the formation time of a giant planet is for practical purposes very close to the time needed to reach the critical mass \citep[see e.g.][]{1986Icar...67..391B, 1996Icar..124...62P, 2005A&A...434..343A}. 

However, this conclusion only holds for large values of the critical core, as the gas accretion rate at the critical mass is a rapidly increasing function of the planetary mass. As we have shown here, the condensation of water reduces drastically the critical mass, and as a consequence, the gas accretion rate at the critical mass could be in this context very small compared to standard formation models. The work presented here shows that the standard picture of planet formation, including the concept of critical \textit{core} mass, is overly simplified and should be revised in the future.

\paragraph{Acknowledgments.}
 We thank C. Mordasini for fruitful discussions and the constructive remarks of an anonymous referee who contributed to improve this paper. We acknowledge financial support from the organising committee of the 1st and 2nd ELSI Symposium and of the  International Space Science Institute, in the framework of an ISSI Team. This work has been in part carried out within the frame of the National Centre for Competence in Research PlanetS supported by the Swiss National Science Foundation. The authors acknowledge the financial support of the SNSF, the European Research Council under grant 239605, and by Grants-in-Aid for Scientific Research on Innovative Areas (No. 23103005) and Scientific Research (C) (No. 25400224) from the Ministry of Education, Culture, Sports, Science and Technology (MEXT) of Japan.

\bibliographystyle{aa}
\bibliography{lit}{}

\end{document}